\documentclass[modern]{aastex61}

\submitjournal{the Astrophysical Journal}

\usepackage{bm}

\newcommand{\normal}[2]{{\cal N} \left( #1, #2 \right)}

\newcommand{\veclambda}{{\bm \lambda}}
\newcommand{\drawnfrom}{\;\sim\;}
\newcommand{\matern}{Mat\'{e}rn}

\newcommand{\kms}{\mathrm{km\;s}^{-1}}
\newcommand{\obj}{LP661-13}

\shorttitle{Disentangling Spectra w/ Gaussian Processes}
\shortauthors{Czekala et al.}

\begin{document}

\title{Disentangling Time Series Spectra with Gaussian Processes: Applications to Radial Velocity Analysis}

\correspondingauthor{Ian Czekala}
\email{iczekala@stanford.edu}

\author[0000-0002-1483-8811]{Ian Czekala}
\altaffiliation{KIPAC Postdoctoral Fellow}
\affiliation{Kavli Institute for Particle Astrophysics and Cosmology,\\
Stanford University, Stanford, CA 94305, USA}

\author[0000-0001-9846-4417]{Kaisey S. Mandel}
\affiliation{Harvard-Smithsonian Center for Astrophysics, \\
60 Garden Street, Cambridge, MA 02138, USA}

\author[0000-0003-2253-2270]{Sean M. Andrews}
\affiliation{Harvard-Smithsonian Center for Astrophysics, \\
60 Garden Street, Cambridge, MA 02138, USA}

\author[0000-0001-7730-2240]{Jason A. Dittmann}
\affiliation{Harvard-Smithsonian Center for Astrophysics, \\
60 Garden Street, Cambridge, MA 02138, USA}

\author[0000-0001-8351-408X]{Sujit K. Ghosh}
\affiliation{Department of Statistics, NC State University,\\
2311 Stinson Dr., Raleigh, NC 27695, USA}
\affiliation{Statistical and Applied Mathematics Institute,\\
19 T.W. Alexander Drive, Research Triangle Park, NC 27709, USA}

\author[0000-0001-7516-8308]{Benjamin T. Montet}
\altaffiliation{NASA Sagan Fellow}
\affiliation{Department of Astronomy and Astrophysics, \\
University of Chicago, \\
5640 S. Ellis Ave., Chicago, IL 60637, USA}

\author[0000-0003-4150-841X]{Elisabeth R. Newton}
\altaffiliation{NSF Astronomy and Astrophysics Postdoctoral Fellow}
\affiliation{Massachusetts Institute of Technology,\\
Cambridge, MA, 02138, USA}

\begin{abstract}

Measurements of radial velocity variations from the spectroscopic monitoring of stars and their companions are essential for a broad swath of astrophysics, providing access to the fundamental physical properties that dictate all phases of stellar evolution and facilitating the quantitative study of planetary systems. The conversion of those measurements into both constraints on the orbital architecture and individual component spectra can be a serious challenge, however, especially for extreme flux ratio systems and observations with relatively low sensitivity.
Gaussian processes define sampling distributions of flexible, continuous functions that are well-motivated for modeling stellar spectra, enabling proficient search for companion lines in time-series spectra. We introduce a new technique for spectral disentangling, where the posterior distributions of the orbital parameters and intrinsic, rest-frame stellar spectra are explored simultaneously without needing to invoke cross-correlation templates.
To demonstrate its potential, this technique is deployed on red-optical time-series spectra of the mid-M dwarf binary LP661-13.  We report orbital parameters with improved precision compared to traditional radial velocity analysis and successfully reconstruct the primary and secondary spectra. We discuss potential applications for other stellar and exoplanet radial velocity techniques and extensions to time-variable spectra. The code used in this analysis is freely available as an open source Python package.

\end{abstract}

\keywords{techniques: radial velocity --- techniques: spectroscopic --- binaries : spectroscopic ---
stars: fundamental parameters --- celestial mechanics --- stars: individual (LP661-13)}

\section{Introduction} \label{sec:intro}

Close binary pairs are the foundation of stellar astrophysics.  Measurements of the orbital dynamics in these systems can constrain masses, the fundamental physical parameter in stellar evolution.  That information is vital in our understanding of everything from star formation to death, with wide-reaching implications for issues ranging from molecular cloud collapse and fragmentation to exoplanets to cosmology with Type Ia supernovae  \citep{torres10}.
A common means of measuring binary orbital dynamics is through high resolution spectroscopic radial velocity monitoring. Traditionally, radial velocities for each stellar component are measured by cross-correlating an observed composite spectrum with various Doppler-shifted stellar templates \citep[e.g., \texttt{TODCOR};][]{zucker94}. The velocity for each component corresponds to the Doppler shift which delivers the maximum cross-correlation signal. While cross-correlation is commonly employed as a  workhorse technique, there are several shortcomings to this limited statistical framework. First, in the case of low signal to noise data, there can be considerate uncertainty about how well the chosen templates match the true spectra of the stars. In its most straightforward form, cross-correlation is unable to meaningfully account for variable spectral lines or uncertain calibration parameters in a principled probabilistic framework, nuisances which can systematically bias a radial velocity measurement. Moreover, at no point does the cross-correlation framework reconstruct either of the component spectra, preventing a check of the suitability of the chosen template as well as any detailed photospheric analysis of the stars themselves.

Because the composite spectra of spectroscopic binaries are modulated by a deterministic Doppler shift, it is possible to ``disentangle'' the intrinsic spectra from multiple observations at different orbital phases. One of the first successful attempts was by \citet{bagnuolo91}: by viewing the composite spectra of the double-lined O star AO Cas as various projections of the intrinsic component spectra, they were able to apply iterative tomographic reconstruction techniques and recover the component spectra. Soon after, \citet{simon94} developed a sparse matrix formalism to decompose composite spectra into their components while also optimizing the orbital elements of the binary star. In contrast to traditional radial velocity techniques, spectral disentangling techniques can determine the intrinsic spectra and velocities simultaneously, simplifying the coherent propagation of uncertainties and often resulting in more precise constraints on the orbital parameters \citep{pavlovski10}. Spectral decomposition can provide precise radial velocities even at orbital phases where the lines from both stars overlap, an area of difficulty for cross-correlation techniques. Once disentangled, the component spectra can be further analyzed with conventional spectroscopic techniques to determine the fundamental stellar properties. For example, \citet{rawls16} disentangled the spectra of the double red giant eclipsing binary KIC~9246715 and then used the radiative transfer code \texttt{MOOG} \citep{sneden73} to estimate the effective temperature, surface gravity, and metallicity for each star.

A major advance in spectral disentangling was the realization that the reconstruction could be performed quickly in the Fourier domain, thanks to the efficiency of the FFT \citep[\texttt{KOREL};][]{hadrava95}. However, the FFT introduces some potentially undesirable side effects. Because the FFT treats the spectrum as a continuous, periodic function, it is important to carefully choose ``chunks'' of the spectrum such that the edges are at the same continuum level, otherwise spectral lines will redshift and blueshift off the edge of the chunk and distort the reconstruction \citep{ilijic04a}. This problem can be particularly acute when dealing with a late spectral type star with a high density of spectral lines.
While Fourier techniques provide the ability to filter out high frequency noise (de-noising), they may also have difficulty constraining low-order ``continuum''-like features of the spectra, making reconstructed spectra appear wavy \citep{hadrava95}. Lastly, the FFT requires that input spectra be interpolated to a log-$\lambda$ spaced wavelength grid (a process that introduces correlated noise) and treats the measurement errors as homoscedastic. This means that if the signal-to-noise of the spectrum is wavelength-dependent or there happens to be a cosmic-ray hit on a certain region of the detector, there is no straightforward way to adjust the weighting of specific portions of the spectrum.

The flexible probabilistic framework offered by Gaussian processes potentially provides a means to address many of the limitations of traditional orbit measurement and spectral disentangling techniques. In truth, stellar spectra are not actually physically generated from Gaussian processes but are rather the result of non-linear radiative transfer through complex stellar atmospheres with specific wavelength-dependent elemental and molecular opacities. However, for the purposes of inferring intrinsic spectra and stellar radial velocities, simple Gaussian processes offer an attractive framework to model stellar spectra in a purely data driven manner. Gaussian processes have been used successfully for other time-series applications in astronomy, for example modeling lensed quasar time delays \citep{hojjati13, tak16}, inferring stellar rotation periods \citep{angus15}, and modeling correlated noise in photometric observations of planet transits and eclipses \citep{evans15, montet16}.

The content of this paper is as follows: in \S\ref{sec:model}, we introduce Gaussian processes and demonstrate how they may be used to model a stellar spectrum. Then, we model a mock double-lined spectroscopic binary--where spectral lines from both components are seen in the composite spectrum--and demonstrate how to simultaneously infer the orbital parameters and the intrinsic spectra of both stars. We also show how the precision of the orbital posteriors and quality of the reconstructed spectra respond to changes in the binary flux ratio and signal-to-noise of the dataset. In \S\ref{sec:results} we apply our technique to the mid-M binary system \obj, recently studied by \citet{dittmann16}. We demonstrate precise inference of the orbital parameters and reconstruct the spectra of A and B, which are an excellent match to other mid-M spectral templates.
In \S\ref{sec:discussion}, we discuss potential extensions of the Gaussian process framework to variable stellar spectra, telluric line modeling, and precision radial velocity measurement for exoplanet detection, and in \S\ref{sec:summary} we conclude the paper.

\section{Gaussian Process Spectral Models} \label{sec:model}
In this section, we describe a framework for modeling a stellar spectrum, and sums of stellar spectra, as Gaussian processes. First, we review common notation and theorems for multidimensional Gaussian random variables. Second, we explain how a Gaussian process can be used to model a spectrum of a single star which is stationary in time and shows no orbital motion due to the presence of a companion. Third, we introduce a Keplerian orbital model and extend the Gaussian process framework to model the spectral time series of a double-lined spectroscopic binary. Throughout this section we use archival observations of a single star to simulate observations and demonstrate the development of the framework.

We adopt the notation that
\begin{equation}
  {\bm x} \drawnfrom \normal{{\bm \mu}_x}{{\bm \Sigma}_{xx}}
  \label{eqn:GPprior}
\end{equation}
signifies that the vector ${\bm x}$ is drawn from a multi-dimensional Gaussian distribution with mean vector ${\bm \mu}$ and covariance matrix ${\bm \Sigma}_{xx}$. The elements of ${\bm \Sigma}_{xx}$ are the covariances between pairwise elements within ${\bm x}$ and can be specified directly or via a functional prescription. By definition, the likelihood function associated with ${\bm x}$ is the multi-dimensional Gaussian
\begin{equation}
  p({\bm x} |\, {\bm \mu}_x, {\bm \Sigma}_{xx}) = \frac{1}{\left [(2 \pi)^N \det{{\bm \Sigma}_{xx}} \right]^{1/2}} \exp \left ( - \frac{1}{2} ({\bm x} - {\bm \mu}_x)^\mathrm{T} {\bm \Sigma}_{xx}^{-1} ({\bm x} - {\bm \mu}_x) \right)
  \label{eqn:GPlikelihood}
\end{equation}
where $N$ is the length of ${\bm x}$. For computational reasons, the natural logarithm of the likelihood is frequently used
\begin{equation}
\ln p({\bm x} |\, {\bm \mu}_x, {\bm \Sigma}_{xx}) = -\frac{1}{2}\left [ ({\bm x} - {\bm \mu}_x)^\mathrm{T} {\bm \Sigma}_{xx}^{-1} ({\bm x} - {\bm \mu}_x) + \ln \det {\bm \Sigma}_{xx} + N \ln 2 \pi \right ]
\label{eqn:GPlnlikelihood}
\end{equation}
Following \citet[][e.q. A.5]{rasmussen05}, we let ${\bm x}$ and ${\bm y}$ be jointly Gaussian random vectors drawn from
\begin{equation}
  \left [
  \begin{array}{c}
  {\bm x} \\
  {\bm y}
  \end{array} \right ]
  \drawnfrom
  \normal{
  \left [
  \begin{array}{c}
  {\bm \mu_x }\\
  {\bm \mu_y }
  \end{array}
  \right ]}{
  \left [
  \begin{array}{cc}
    {\bm \Sigma}_{xx} & {\bm \Sigma}_{xy} \\
    {\bm \Sigma}_{yx} & {\bm \Sigma}_{yy}
  \end{array}
  \right ]},
  \label{eqn:joint}
\end{equation}
where ${\bm \mu}_x$ and ${\bm \mu}_y$ are the vector means. The sub-matrices ${\bm \Sigma}_{xx}$ and ${\bm \Sigma}_{yy}$ are the covariance matrices corresponding to the elements within ${\bm x}$ and ${\bm y}$, and ${\bm \Sigma}_{xy}$ encapsulates the covariances between the elements in ${\bm x}$ and ${\bm y}$; ${\bm \Sigma}_{yx} = {\bm \Sigma}_{xy}^\mathrm{T}$.
If and only if ${\bm x}$ and ${\bm y}$ are independent will ${\bm \Sigma}_{xy} = {\bm 0}$. The conditional distribution of ${\bm x}$ given ${\bm y}$ is also a Gaussian distribution
\begin{equation}
   {\bm x} | {\bm y} \drawnfrom
   \normal{{\bm \mu_x } + {\bm \Sigma}_{xy}{\bm \Sigma}_{yy}^{-1} ({\bm y} - {\bm \mu_y})}{
   {\bm \Sigma}_{xx} - {\bm \Sigma}_{xy}{\bm \Sigma}_{yy}^{-1}{\bm \Sigma}_{yx}}.
   \label{eqn:conditional}
\end{equation}
These equations are the foundation for constructing the Gaussian process spectra model that follows.

\subsection{A model for observations of a single star}
For our spectroscopic application, the input vector is the sampled wavelengths of the detector $\{ \lambda_i\}_{i=1}^w$ and the data vector is the observed fluxes $\{d_i\}_{i=1}^w$
\begin{eqnarray}
  {\bm \lambda} =
  \left [
  \begin{array}{c}
    \lambda_1 \\
    \lambda_2 \\
    \vdots \\
    \lambda_w \\
  \end{array}
  \right ] \qquad
  {\bm d} = \left [
  \begin{array}{c}
    d_1 \\
    d_2 \\
    \vdots \\
    d_w \\
  \end{array}
  \right]
\end{eqnarray}
where each pixel is indexed by $i$ from $1$ to $w$, the number of pixels in the region of the spectrum under consideration. To illustrate the development of our framework, we use synthetic ``mock'' datasets generated by the following recipe. First, we create a template by stacking many archival high-resolution observations of the K5 star LkCa~14 and smoothing the result with a Gaussian kernel. Then, we sample the spectrum at ${\bm \lambda}$ and add a known amount of independent white noise to the dataset.

We will model the continuous, intrinsic stellar spectrum as a Gaussian process. A function $f(\lambda)$, where $\lambda > 0$, is said to have a Gaussian process with mean function $\mu(t)$ and covariance kernel $k(\lambda, \lambda^\prime)$ if for any finite collection of inputs $0 < \lambda_1 < \lambda_2 < \ldots < \lambda_w$
the vector $\{f(\lambda_1),\, f(\lambda_2),\, \ldots,\, f(\lambda_w)\}$ has a multivariate Gaussian distribution with mean vector $\{\mu(\lambda_1),\, \mu(\lambda_2),\, \ldots,\, \mu(\lambda_w) \}$ and $w \times w$ covariance matrix with elements $k(\lambda_i, \lambda_j)$, where $i, j = 1, 2, \ldots, w$, and $k$ is a positive definite kernel function.
A single function generated from a Gaussian process is called a realization of the Gaussian process. We consider the continuous, intrinsic stellar spectrum $f$ to be a function generated from the Gaussian process
\begin{equation}
  f(\lambda) \drawnfrom \mathrm{GP}(\, \mu(\lambda), k(\lambda, \lambda^\prime)).
  \label{eqn:GPdrawnfrom}
\end{equation}
For the purposes of this paper, we will always work with finite-length samplings of the Gaussian process ${\bm f}$, either at the detector wavelengths $\veclambda$ or finely spaced vectors of our own design.
To model a single epoch of a single star, we treat the data ${\bm d}$ as the sum of a realization ${\bm f}$ of the Gaussian process with a realization of the noise process ${\bm N}$,
\begin{eqnarray}
  {\bm d}\, = & {\bm f} + {\bm N} \\
  {\bm d} \drawnfrom & \normal{{\bm \mu}_f}{ {\bm \Sigma}_f} + \normal{{\bm 0}}{ {\bm \Sigma}_N} \\
  {\bm d} \drawnfrom & \normal{{\bm \mu}_f}{ {\bm \Sigma}_f + {\bm \Sigma}_N}
\end{eqnarray}
where ${\bm \Sigma}_N$ is a covariance matrix describing the noise of the dataset. For most spectroscopic observations, this will simply be a diagonal matrix with the Poisson uncertainty for each pixel, unless a particular telescope reduction pipeline provides additional information about inter-pixel covariances. It is assumed that the spectrum has been rectified to have a mean flux of 1, and so by design we choose the Gaussian process to have ${\bm \mu}_f = {\bm 1}$. In theory, we could allow the mean to vary but in practice it has little effect on the result as long as a reasonable number (or function of $\lambda$) is chosen to be $\sim 1$. The covariance matrix ${\bm \Sigma}_f$ is populated by evaluating a kernel function between pairs of input values in $\veclambda$. The choice of the kernel function determines the smoothness of the functions drawn from the Gaussian process. As a distance metric, we compute the velocity distance between two pairs of wavelengths $i$ and $j$ as
\begin{equation}
  r_{ij} = r(\lambda_i, \lambda_j) = \frac{c}{2} \left | \frac{\lambda_i - \lambda_j}{\lambda_i + \lambda_j} \right |,
\end{equation}
where $c$ is the speed of light and $r_{ij}$ has units of $\kms$. A popular kernel is the squared exponential kernel, which we choose to model our spectra. This kernel specifies the covariance between two pixels $\lambda_i$ and $\lambda_j$ as
\begin{equation}
  k_{ij}(r_{ij} |\, a, l) = a^2 \exp \left ( -\frac{r_{ij}^2}{2 l^2}\right),
  \label{eqn:kernel}
\end{equation}
 where $a$ sets the amplitude of the Gaussian process and has the same units as the rectified flux, and $l$ sets the length scale of the Gaussian process and has units of $\kms$. In this paper, we refer to $a$ and $l$ as Gaussian processes hyperparameters.

\begin{figure}[htb]
\begin{center}
 \includegraphics{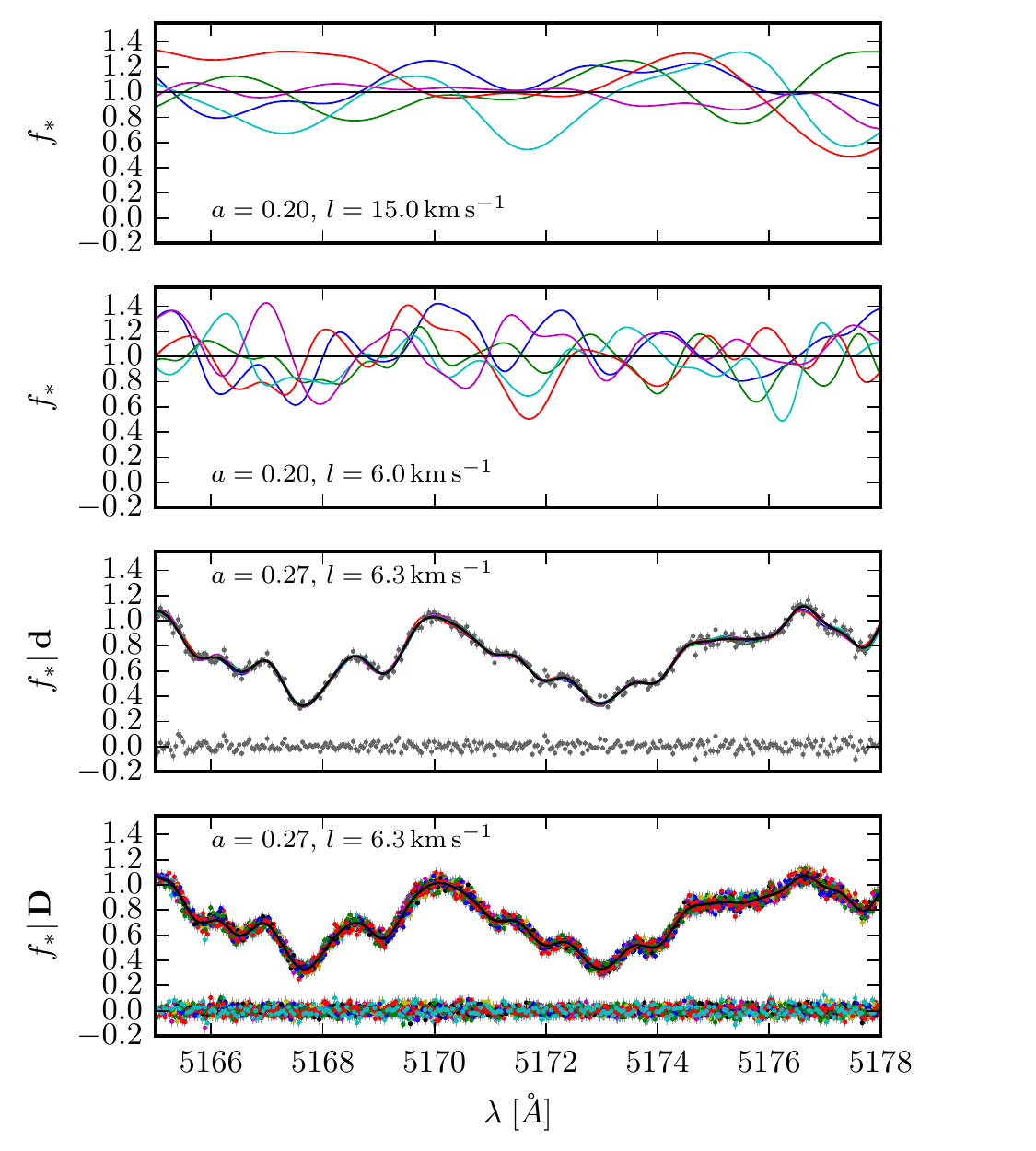}
  \figcaption{\emph{top two panels}: multivariate draws from the prior distribution of functions, given by Equation~\ref{eqn:GPprior_predict} using a mean of ${\bm 1}$ (black line) and the Gaussian process hyperparameters $a$ and $l$ specified in the figure. Completely unconstrained by data, the possible realizations of the Gaussian process span a large range of flux space. As the length scale $l$ is decreased, the correlation length of the realizations also shortens.  \emph{third panel}: draws of the Gaussian process conditioned on the observed data from a single spectroscopic observation, with the maximum likelihood estimates. Multiple draws from the posterior predictive distribution are shown, with the mean prediction in black and the residuals at the bottom, showing that a Gaussian process can be used to accurately model spectra.
   \emph{bottom panel}: Multiple observations of the same star, taken over a series of nights. The sampling density increases dramatically due to the shifting of the barycentric frame.
  \label{fig:single_star}}
  \end{center}
\end{figure}

Now we have fully specified the components of the sampling distribution for the Gaussian process $f$ (Equation~\ref{eqn:GPprior}), which can be thought of as a distribution over functions. We use the ${}_*$ subscript to denote realizations of a vector, which is a draw from this distribution
\begin{equation}
  {\bm f}_* \drawnfrom \normal{{\bm 1}}{{\bm \Sigma}_f + {\bm \Sigma}_N}.
  \label{eqn:GPprior_predict}
\end{equation}
Several realizations of ${\bm f}_*$ are shown in the top panels of Figure~\ref{fig:single_star}, for two different choices of the Gaussian process hyperparameters $a$ and $l$. While the Gaussian process has a mean of 1, actual realizations of the Gaussian process scatter about the mean. As $l$ decreases, the covariance between distant elements decreases and ${\bm f}_*$ oscillates more rapidly.
How does one choose the best values of $a$ and $l$? Fortunately, the Gaussian process framework provides a natural mechanism to determine the most probable values through the Gaussian process likelihood, given by Equation~\ref{eqn:GPlikelihood}.\footnote{We will use Equations~\ref{eqn:GPlikelihood},~\ref{eqn:joint},~\&~\ref{eqn:conditional} throughout this paper and the values of ${\bm \Sigma}_{xx}$, ${\bm \Sigma}_{yy}$, and ${\bm \Sigma}_{xy}$ will change depending on the context.}
In the presence of data (${\bm x} = {\bm d}$), a choice of mean vector (${\bm \mu}_x = {\bm 1}$), a covariance matrix (${\bm \Sigma}_{xx} = {\bm \Sigma}_f + {\bm \Sigma}_N$), and flat priors on the hyperparameters, the posterior probability distribution is directly specified by the marginal likelihood, Equation~\ref{eqn:GPlikelihood}, which we can maximize with respect to $a$ and $l$.

With the hyperparameters optimized, we can now make realizations of the Gaussian process conditioned on the data by taking random draws from the posterior predictive distribution (Equation~\ref{eqn:conditional}). These draws represent realizations of the inferred intrinsic stellar spectrum while the scatter in the draws serves to illustrate the uncertainty in our inference.
In this application of Equation~\ref{eqn:conditional},  ${\bm x} = {\bm f}_\ast$; ${\bm y} = {\bm d}$; ${\bm \Sigma}_{xx}$ is filled out by the kernel evaluated over the prediction wavelengths ${\bm \lambda}_\ast$ corresponding to ${\bm f}_\ast$; ${\bm \Sigma}_{yy} = {\bm \Sigma}_f + {\bm \Sigma}_N$, where  ${\bm \Sigma}_f$ is filled out by the kernel evaluated over the wavelengths ${\bm \lambda}$ corresponding to ${\bm d}$; and ${\bm \Sigma}_{xy}$ is filled out by the kernel evaluated over pairs of wavelengths corresponding to both ${\bm \lambda}_\ast$ and ${\bm \lambda}$.
Random draws of the Gaussian process ``snap'' to the data (Figure~\ref{fig:single_star}, third panel), providing a very flexible mechanism for modeling the continuous stellar spectrum.

Although the Gaussian process achieves promising success in this single-observation application, its real benefit accrues when there are multiple observations of the same star in a time series. Throughout this paper, we assume that all spectroscopic observations of a target star are acquired by the same telescope, and that the instrument line spread function (LSF) is stable across epochs. For the applications discussed in this paper, these criterion are satisfied by most high-resolution spectrographs, such as echelle spectrographs used for radial velocity planet-searches. Because convolution is a linear operation, it is reasonable to ignore the effects of the line spread function and assume that the intrinsic stellar spectrum we are modeling has already been convolved. For very precise applications (e.g. radial velocity search for earth-mass planets), variations in the line spread function may need to be considered; we discuss this further in \S\ref{subsec:hierarchical}.

\begin{figure}[t]
 \begin{center}
  \includegraphics{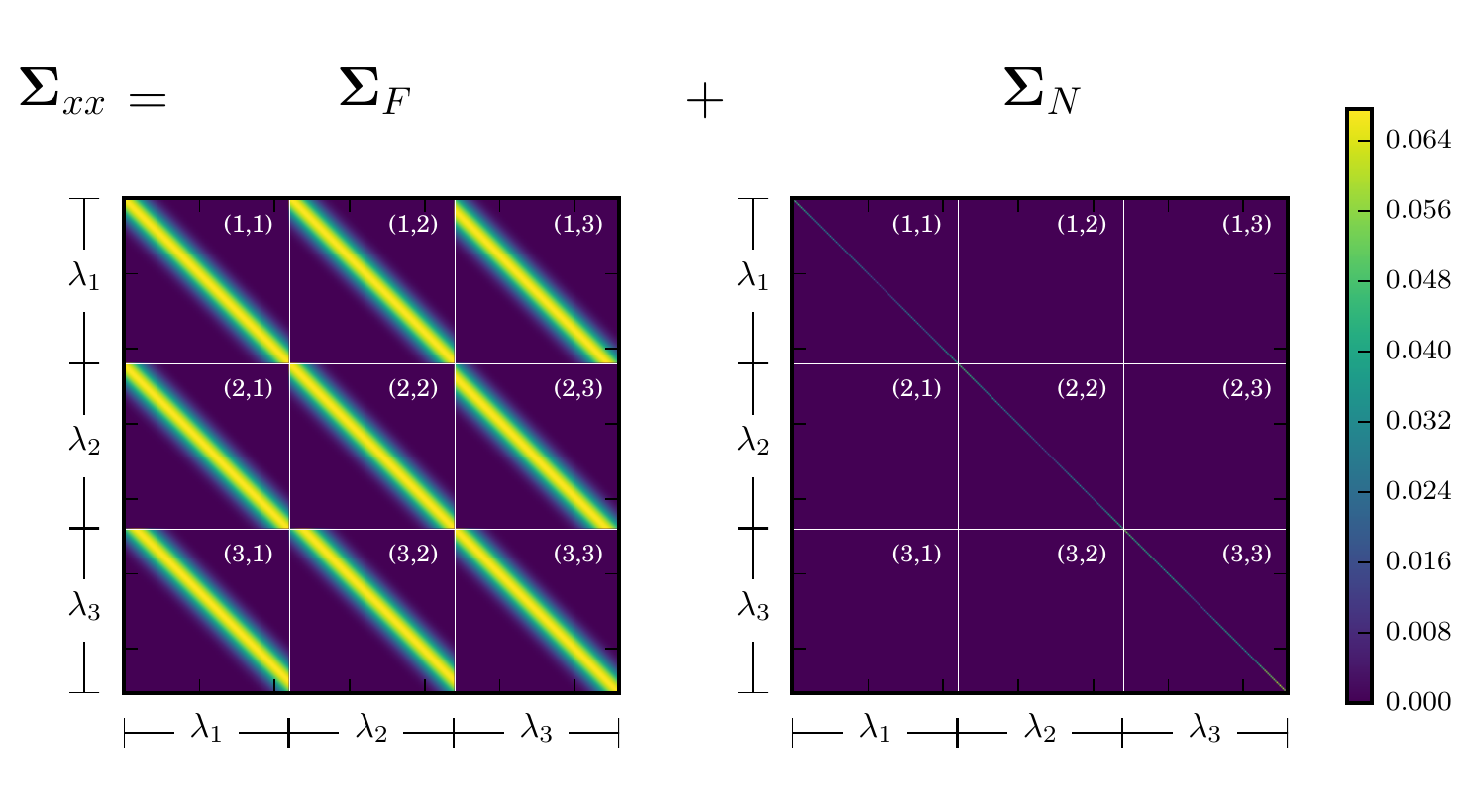}
   \figcaption{The covariance matrix ${\bm \Sigma}_{xx} = {\bm \Sigma}_F + {\bm \Sigma}_N$ for multiple epoch observations of the same star. Off-diagonal sub-matrices ($\{ {\bm \lambda}_k, {\bm \lambda}_l \}$, $k \ne l$) signify the covariances between nearby wavelengths that are separated in time across epochs. The bands of covariances in the off-diagonal blocks are not exactly centered along the block diagonals because the different barycentric corrections for each epoch mean that the wavelength samplings are different for each epoch. In this representation, ${\bm \Sigma}_N$ has been multiplied by 50 to better illustrate the structure of the matrix. The elements of ${\bm \Sigma}_{xx}$ have units of flux squared (rectified to 1).
   \label{fig:V11_multi_epoch}}
   \end{center}
 \end{figure}

Even though spectrographs may be very stable, such that from epoch to epoch nearly the same wavelengths are sampled by the detector in the topocentric frame, this is not the case in the \emph{barycentric frame}. Due to the rotation of the earth and its orbital motion around the sun, observations taken of a source on different days will likely have different relative velocities between the topocentric frame and the barycentric frame, the so-called barycentric correction. This means that at each epoch, the pixels of the detector are sampling slightly different wavelengths of the intrinsic stellar spectrum (broadened by the LSF). This barycentric correction can be computed to high accuracy \citep[e.g.,][]{wright14}, and when all spectra in a time series are corrected to the barycentric frame we have a very dense (albeit noisy) sampling of the convolved stellar spectrum (see Figure~\ref{fig:single_star}, bottom panel). We concatenate the $W$ epochs of data and denote new variables
 \begin{eqnarray}
   {\bm \Lambda} =
   \left [
   \begin{array}{c}
     {\bm \lambda}_1 \\
     {\bm \lambda}_2 \\
     \vdots \\
     {\bm \lambda}_W \\
   \end{array}
   \right  ], \qquad
   {\bm D} =
   \left [
   \begin{array}{c}
     {\bm d}_1 \\
     {\bm d}_2 \\
     \vdots \\
     {\bm d}_W \\
   \end{array}
   \right  ], \qquad
   {\bm F} =
   \left [
   \begin{array}{c}
     {\bm f_1} \\
     {\bm f_2} \\
     \vdots \\
     {\bm f_W} \\
   \end{array}
   \right  ]
 \end{eqnarray}
 and
 \begin{equation}
   {\bm D} = {\bm F} + {\bm N}.
 \end{equation}
The aforementioned covariance matrices are generated from the same kernel as before, with the major difference that it is now applied to all of the input wavelengths in all epochs, which creates a covariance matrix structure seen in Figure~\ref{fig:V11_multi_epoch}.  In principle, we could sort the data in increasing wavelength to create a covariance matrix with a single, central diagonal band; instead we choose to keep adjacent wavelengths in the same epoch next to each other for later purposes. Posterior predictive draws for the multi-epoch Gaussian process are shown in the bottom panel of Figure~\ref{fig:single_star}.

\clearpage

\subsection{A model for observations of a spectroscopic binary} \label{subsec:orbital_params}
Now, we extend the Gaussian process model to include multiple, time series observations of a spectroscopic binary star system. Typically, labeling a system a spectroscopic binary implies that spectral lines from both stars are seen in the composite spectra, however we will later show that a simplified version of this framework can also be used to model single-lined spectroscopic binary systems (e.g., stars hosting exoplanets). The orbital motion of each star induces a Doppler shift of the rest-frame stellar spectra, which are then simultaneously observed in composite. If a star is moving with a radial velocity $v$ relative to the barycentric frame, the rest-frame wavelengths $\veclambda_0$ of the spectrum are shifted to
\begin{equation}
  \veclambda(v) = \sqrt{\frac{c + v}{c - v}}\; \veclambda_0
  \label{eqn:Doppler}
\end{equation}
where a positive $v$ denotes a redshift, or increase in the values of the rest frame wavelengths.
The radial velocities of binary stars as a function of time can be fully described by seven parameters $\theta$ = \{$q$, $K_A$, $e$, $\omega$, $P$, $T_0$, $\gamma$ \}, the mass ratio $q = M_B/ M_A = K_A/K_B$, the velocity semi-amplitude of the primary star, the eccentricity, argument of periastron, orbital period, epoch of periastron, and systemic velocity, respectively \citep{murray10}. The mean anomaly $M$ is given by Kepler's equation
\begin{equation}
  M(t) = E(t) - e \sin E(t) = \frac{2 \pi t - T_0}{P}
\end{equation}
which must be solved to find the eccentric anomaly $E$. The true anomaly $f$ is given by
\begin{equation}
  \cos f(t) = \frac{\cos E(t) - e}{1 - e \cos E(t)}.
\end{equation}
Then, the velocity of the primary star as a function of time is
\begin{equation}
  v_A = K_A \left (\cos (\omega + f(t)) + e \cos \omega \right ) + \gamma
\end{equation}
and the secondary velocity is
\begin{equation}
  v_B = -\frac{K_A}{q} \left (\cos (\omega + f(t)) + e \cos \omega \right ) + \gamma.
\end{equation}
The resulting spectroscopic binary dataset has the same dimensionality as the single star dataset, but we now assume that it is produced from the sum of the spectrum of the primary star, denoted by a realization $f$ of a Gaussian process, and the spectrum of the secondary star, denoted by a realization $g$ of an (independent) Gaussian process. The secondary spectrum is produced from a Gaussian process similar to $f$ (e.g., Equation~\ref{eqn:GPdrawnfrom}) with the same form of covariance kernel (Equation~\ref{eqn:kernel}). We denote the various components of the spectroscopic binary dataset as
\begin{equation}
  {\bm \Lambda} =
  \left [
  \begin{array}{c}
    {\bm \lambda}_1 \\
    {\bm \lambda}_2 \\
    \vdots \\
    {\bm \lambda}_W \\
  \end{array}
  \right  ], \qquad
  {\bm D} =
  \left [
  \begin{array}{c}
    {\bm d_1} \\
    {\bm d_2} \\
    \vdots \\
    {\bm d_W} \\
  \end{array}
  \right  ], \qquad
  {\bm F} =
  \left [
  \begin{array}{c}
    {\bm f_1} \\
    {\bm f_2} \\
    \vdots \\
    {\bm f_W} \\
  \end{array}
  \right  ], \qquad
  {\bm G} =
  \left [
  \begin{array}{c}
    {\bm g_1} \\
    {\bm g_2} \\
    \vdots \\
    {\bm g_W} \\
  \end{array}
  \right  ]
\end{equation}
\begin{equation}
  {\bm D} = {\bm F} + {\bm G} + {\bm N}.
\end{equation}
This spectroscopic binary application can be thought of as modeling the rest-frame spectrum of the primary star (star A), and separately that of the secondary star (star B), while only having access to datasets which represent the sum of the spectra at different orbital phases, all in the presence of noise.  We are assuming that the two Gaussian processes are independent from each other and that the Doppler shifts from the orbital motion will allow us to disentangle them. For demonstration purposes, we assume that the orbital parameters $\theta$ are known, and therefore we can calculate the radial velocities of the primary and secondary at all epochs. To fill out the covariance matrix, we must first create new input vectors corresponding to $f$ and $g$ that relate the observed flux at each epoch to shifted versions of the rest-frame spectra of A and B,
\begin{equation}
  {\bm \Lambda}_F =
  \left [
  \begin{array}{c}
    {\bm \lambda}_{1,A} \\
    {\bm \lambda}_{2,A} \\
    \vdots \\
    {\bm \lambda}_{W,A} \\
  \end{array}
  \right  ], \qquad
  {\bm \Lambda}_G =
  \left [
  \begin{array}{c}
    {\bm \lambda}_{1,B} \\
    {\bm \lambda}_{2,B} \\
    \vdots \\
    {\bm \lambda}_{W,B} \\
  \end{array}
  \right  ].
\end{equation}
To produce these vectors, the original sub-components of ${\bm \Lambda}$ are shifted by the predicted radial velocity for each star at each epoch according to Equation~\ref{eqn:Doppler}, such that ${\bm \Lambda}_F$ and ${\bm \Lambda}_G$ should correspond to the wavelengths of star A and B in their respective rest-frames. To be explicit with notation, these input vectors are actually a function of the orbital parameters, ${\bm \Lambda}_F (\theta)$ and ${\bm \Lambda}_G (\theta)$.

\begin{figure}[t]
\begin{center}
 \includegraphics{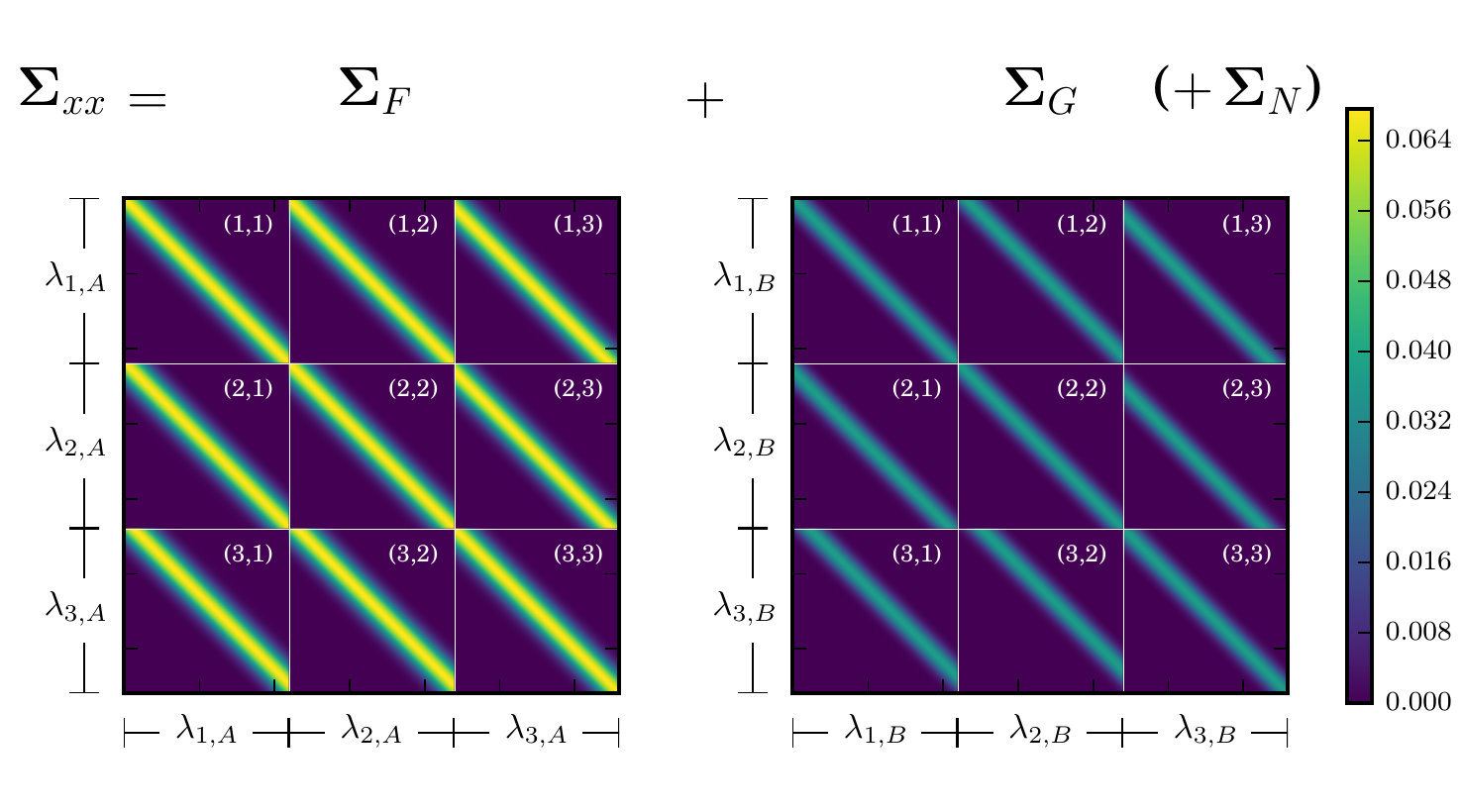}
  \figcaption{For multiple observations of a spectroscopic binary, the structure of ${\bm \Sigma}_{xx}$ changes to include the addition of ${\bm \Sigma}_G$. Now, there are separate input wavelength vectors for each matrix, created by Doppler shifting the original wavelength vectors to the rest-frame velocities of each star at that epoch. In this figure, the covariance kernel for the B star is 75\% of the amplitude and length scale for that of A (i.e., $a_g = 0.75 a_f$, $l_g = 0.75 l_f$).
  \label{fig:V11_multi_epoch_binary}}
  \end{center}
\end{figure}

The likelihood function for the sum of $f$ and $g$ is still given by Equation~\ref{eqn:GPlikelihood}, however the covariance matrix is now a composite of two covariance matrices plus the noise matrix,
\begin{equation}
  {\bm \Sigma}_{xx} = {\bm \Sigma}_F + {\bm \Sigma}_G + {\bm \Sigma}_N,
\end{equation}
where ${\bm \Sigma}_F$ is evaluated using ${\bm \Lambda}_F$ and ${\bm \Sigma}_G$ is evaluated using ${\bm \Lambda}_G$ with the kernel specified in Equation~\ref{eqn:kernel}. We show how ${\bm \Sigma}_{xx}$ is constructed in Figure~\ref{fig:V11_multi_epoch_binary}. The off-diagonal submatrices of ${\bm \Sigma}_F$ and ${\bm \Sigma}_G$ denote the cross-epoch covariances of each Gaussian process. Now that we have completely specified the likelihood function for disentangling the component spectra of a spectroscopic binary, we relax the assumption that $\theta$ is fixed. For appropriate values of $a$ and $l$, the likelihood function will be maximized when $\theta$ is chosen such that the Keplerian orbit delivers velocity shifts for each epoch that make the cross-matrices consistently describe the rest-frame spectrum of each star.

With the addition of reasonable priors, we can also explore the full posterior distribution of the orbital parameters $\theta$ and the Gaussian process hyperparameters. If we so desire, we can allow separate values of $a$, $l$ for each spectrum (i.e., $\{a, l \}_f$, $\{a, l\}_g$) or enforce that the Gaussian process hyperparameters are the same. If the spectra of A and B are drastically different in morphology (e.g. an A-type star coupled with an M-type star, or a rapidly rotating primary coupled with a slowly rotating secondary), using separate hyperparameters will yield a more optimal spectral reconstruction. We can also relax the assumption that the Gaussian process hyperparameters are the same at all wavelengths, for example to adapt to a star where the spectral line density changes significantly across the observed wavelength range, which we discuss further in \ref{sec:discussion}.

Next, we focus on the formalism necessary to reconstruct the rest-frame spectra of A and B at a vector of rest-frame wavelengths ${\bm \lambda}_\ast$. The joint probability distribution of the (independent) latent Gaussian processes and the dataset is given by
\begin{equation}
  \left [
  \begin{array}{c}
  {\bm f}_\ast \\
  {\bm g}_\ast \\
  {\bm D} \\
\end{array}
  \right ]
  \drawnfrom
  \normal{
  \left [
  \begin{array}{c}
    {\bm \mu}_f \\
    {\bm \mu}_g \\
    {\bm \mu}_{FG} \\
  \end{array}
  \right ]
  }{
  \left [
  \begin{array}{c@{\hspace{2em}}c@{\hspace{2em}}c}
    {\bm \Sigma}_f & {\bm 0} & {\bm \Sigma}_{fF} \\
    {\bm 0} & {\bm \Sigma}_{g} & {\bm \Sigma}_{gG} \\
    {\bm \Sigma}_{Ff} & {\bm \Sigma}_{Gg} & ({\bm \Sigma}_F + {\bm \Sigma}_G + {\bm \Sigma}_N)\\
  \end{array}
  \right ] },
\end{equation}
where ${\bm \Sigma}_f$ and ${\bm \Sigma}_g$ are evaluated over pairs of wavelengths in ${\bm \lambda}_\ast$; ${\bm \Sigma}_F$ and ${\bm \Sigma}_G$ are evaluated over pairs of wavelengths in ${\bm \Lambda}_F$ and ${\bm \Lambda}_G$, respectively;
and ${\bm \Sigma}_{fF}$ and ${\bm \Sigma}_{gG}$ are evaluated over cross-pairs of wavelengths in (${\bm \lambda}_\ast$, ${\bm \Lambda}_F$) and (${\bm \lambda}_\ast$, ${\bm \Lambda}_G$), respectively. The joint posterior predictive distribution is given by
\begin{equation}
  \left . \left [ \begin{array}{c}
  {\bm f}_\ast \\
  {\bm g}_\ast \\
\end{array}
  \right ] \right |
  {\bm D}
  \drawnfrom
  \normal{
  \left [
  \begin{array}{c}
    {\bm \mu}_f \\
    {\bm \mu}_g \\
  \end{array}
  \right ]
  +
  {\bm \Sigma}_{xy} {\bm \Sigma}_{yy}^{-1} ({\bm D} - {\bm \mu}_{FG})
  }{
  {\bm \Sigma}_{xx} - {\bm \Sigma}_{xy} {\bm \Sigma}_{yy}^{-1} {\bm \Sigma}_{yx}
  }
  \label{eqn:SB2_predictive}
\end{equation}
where
\begin{equation}
  {\bm \Sigma}_{xx} = \left [
  \begin{array}{cc}
    {\bm \Sigma}_f & {\bm 0} \\
    {\bm 0} & {\bm \Sigma}_g \\
  \end{array}
  \right  ],
  \quad\quad
  {\bm \Sigma}_{yy} = {\bm \Sigma}_F + {\bm \Sigma}_G + {\bm \Sigma}_N,
  \quad\quad
  {\bm \Sigma}_{xy} = \left [
  \begin{array}{c}
    {\bm \Sigma}_{fF} \\
    {\bm \Sigma}_{gG} \\
  \end{array}
  \right ].
\end{equation}

While we will soon demonstrate the many advantages of this Gaussian process formalism for disentangling spectra, it also inherits a few of the limitations common to most disentangling techniques. First, there is the somewhat obvious limitation that one needs to observe the binary system at multiple orbital phases to obtain different morphologies of composite spectra. If the orbital period of the system is very long, or by chance all observations were obtained at or near the same orbital phase, we would not be able to separate the composite spectra into individual components. Second, because the intrinsic spectra are modeled nonparametrically, without reference to physical models of spectra, reconstruction techniques are unable to say anything about the stellar velocities in an absolute sense. This means that from spectroscopic reconstruction alone, we are unable to constrain the systemic velocity of the system, $\gamma$, and so this parameter remains fixed to zero in our analysis. Once the spectra are reconstructed, however, $\gamma$ can be easily recovered by cross-correlation with physical spectral templates. Of related consequence is that the reconstruction technique is only able to constrain the relative velocities of each star from epoch to epoch; it is not possible to measure the velocity of star B relative to star A at any epoch with the reconstruction technique alone. However this can be later measured easily by correlation with physical spectra models, or if the two stars in question are bound, inferred through their Keplerian orbits.

\subsection{Application to mock data}
\label{subsec:mock}

\begin{figure}[t]
\begin{center}
 \includegraphics{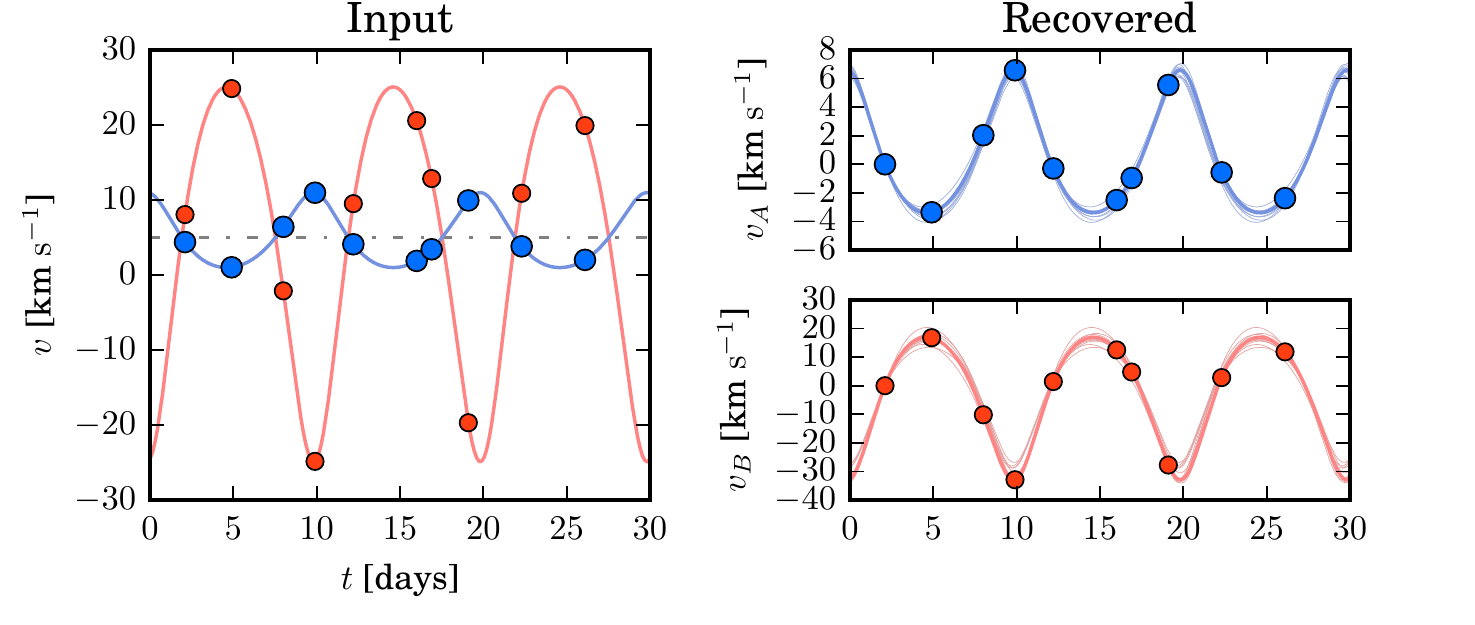}
  \figcaption{\emph{left}: the fiducial orbit for a mock double-lined spectroscopic binary (blue = primary, red = secondary), showing the 10 epochs at which the orbit is sampled. \emph{right}: the recovered orbits for the primary and secondary stars using the $f_B/f_A = 0.2$, SNR = 60 mock dataset. Because the spectral disentangling routines are unable to constrain $\gamma$, the recovered orbits are plotted relative to the velocity of that star at the first observational epoch.
  \label{fig:SB2_orbit}}
  \end{center}
\end{figure}

\begin{figure}[htb]
\begin{center}
 \includegraphics{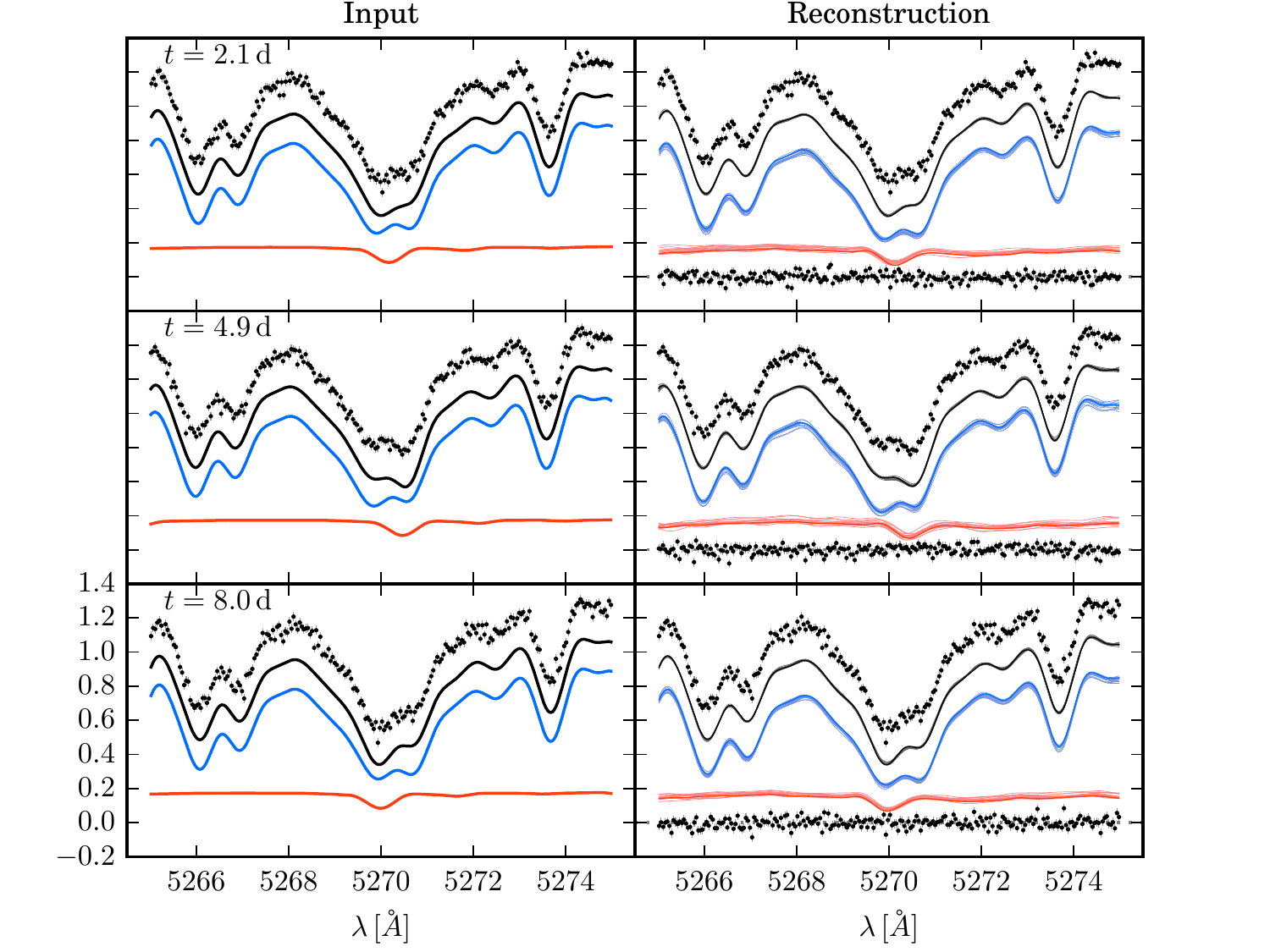}
 \includegraphics{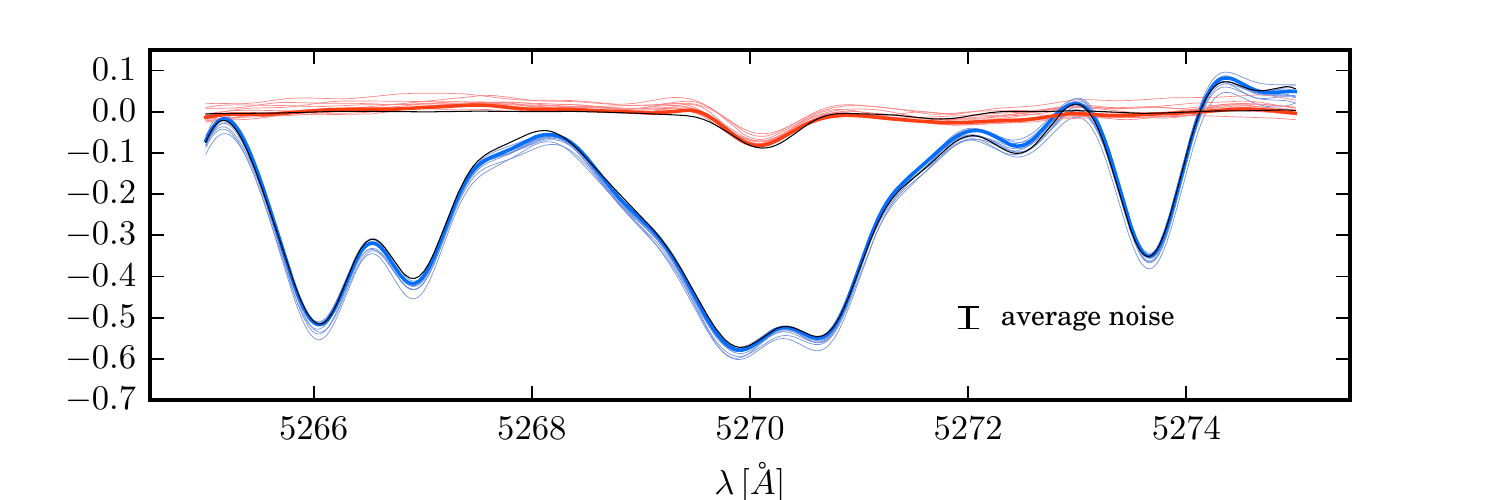}
  \figcaption{\emph{left column}: to generate a mock observation, the primary (blue) and secondary (red) spectra are Doppler-shifted according to the radial velocities calculated from the fiducial orbit, summed to generate the composite spectra (black), and then noise is added. The resulting dataset is shown vertically offset for clarity (black points with errorbars). The full mock dataset consists of 10 observations of a $f_B/f_A = 0.2$ binary with SNR = 60. \emph{right column}: the distribution of the inferred spectra, marginalized over the posterior distributions of the orbital parameters and Gaussian process hyperparameters. The primary and secondary component spectra have been offset by arbitrary flux constants, chosen to correspond to the input flux levels for aesthetic purposes. \emph{bottom}: the input spectra (black) compared against the draws of reconstructed spectra. In this panel, the reconstructed spectra are drawn from a mean zero Gaussian process (${\bm \mu}_f = {\bm \mu}_g = {\bm 0}$) to emphasize that our technique cannot constrain the relative flux level of the components, and so the input spectra have been shifted vertically to match the level of the reconstructed spectra. Because the reconstruction inference leverages multiple epochs of data, we recover the spectra to a significantly higher precision than the average noise in the composite dataset.
  \label{fig:SB2_reconstructed}}
  \end{center}
\end{figure}

To explore the performance of the Gaussian process spectral model, we generate mock spectroscopic binary datasets from a fiducial orbit and assess the accuracy of the recovered stellar spectra and orbital parameters. To generate the datasets, we use segments of real data of the K5 star LkCa~14 observed with the \emph{CfA/TRES} spectrograph on Mt. Hopkins (obtained for a separate purpose under a program with P.I.~I.~Czekala). Multiple epochs of data were combined and smoothed to create a high signal-to-noise template spectrum. We chose two separate regions of this template with an average spectral line density ($5235\AA$ to $5285\AA$ for the primary and $6420\AA$ to $6480\AA$ for the secondary) to mimic spectra from two different stars, and then assigned the secondary fluxes to the same wavelength range as the primary.
We assume a fiducial binary orbit with $\theta = \{K_\mathrm{A} = 5.0\,\kms,\, q = 0.2,\, e = 0.2,\, \omega = 10.0^\circ,\, P = 10.0\,\mathrm{days},\, T_0 = 0.0\,\mathrm{JD},\, \gamma = 5.0\,\kms\}$, sampled at 10 epochs over the course of a $\sim 25\,$day period (see Figure~\ref{fig:SB2_orbit}, left panel).
To generate a composite spectrum for a single epoch, we first Doppler shift each template by its calculated radial velocity, interpolate the secondary spectrum to the same wavelength vector as the primary, multiply each component spectrum by its chosen fractional flux contribution, truncate each template to the range $5265\AA$ to $5275\AA$, sum the spectra, and then add random Gaussian noise according to a chosen signal-to-noise ratio.
This process is illustrated for the first three epochs in the left panel of Figure~\ref{fig:SB2_reconstructed}. While this mock dataset is a small fraction ($< 1/500$) of the temporal and wavelength coverage of a typical radial velocity dataset, it will serve to illustrate the salient characteristics of our technique in a compact and expedient manner.

We first assess the performance of the technique on a mock dataset constructed with moderate flux ratio components and medium-high signal-to-noise ``observations''  ($f_B/f_A = 0.2$, SNR = 60 per resolution element; linespread function FWHM = 2.5 pixels), and then explore the robustness of the technique as these quantities are varied to more extreme values. The posterior distribution of the orbital parameters and Gaussian process hyperparameters is explored simultaneously using a simple implementation of Metropolis-Hastings Markov Chain Monte Carlo (MCMC). The inferred family of orbits is shown in the right panel of Figure~\ref{fig:SB2_orbit}, and these demonstrate good agreement with the input radial velocities. Several draws of reconstructed component spectra and the predicted composite spectrum are shown in the right panel of Figure~\ref{fig:SB2_reconstructed}. To explore the full distribution of inferred component spectra, for each sample we first randomly draw orbital parameters and Gaussian process hyperparameters from the posterior distribution, then draw a sample from the Gaussian process posterior predictive function (Equation~\ref{eqn:SB2_predictive}).

A common limitation of most spectral disentangling techniques is the inability to constrain the continuum level of each star without comparison to a physical model, which our technique suffers from as well. This means that while the relative amplitudes of the variation in each spectrum are well-constrained, there is no knowledge about which star is brighter, i.e., there is a degeneracy between a very luminous companion star with shallow lines and a very faint companion with very deep spectral lines \citep{bagnuolo92}.
As with the $\gamma$ degeneracy, the flux ratio degeneracy can be easily broken by later comparison with a flux-calibrated, physical stellar model \citep{pavlovski10}. In the right columns of Figure~\ref{fig:SB2_reconstructed}, we have vertically offset each recovered spectrum by an arbitrary constant to look similar to the left panels, which has the potential to be misleading. Therefore in the bottom panel of Figure~\ref{fig:SB2_reconstructed}, we show the draws of the reconstructed spectra from mean zero Gaussian processes (${\bm \mu}_f = {\bm \mu}_g = {\bm 0}$) to emphasize that they contain no information about the continuum level. Rather, the input spectra have been subtracted by an arbitrary constant to match the level of the reconstructed spectra.
The reconstructed spectra do an excellent job of reproducing the shape of the input spectra to a precision far surpassing the average pixel noise of the mock dataset. Interestingly, there is a somewhat appreciable scatter in the vertical offset level of the Gaussian process draws.
This is because ${\bm f}_\ast$ and ${\bm g}_\ast$ are drawn from a joint posterior predictive distribution while only their sum is constrained by the data, allowing the mean levels to trade off somewhat (i.e., if a primary draw is slightly higher, then the secondary draw is slightly lower).

\begin{figure}[tb]
\begin{center}
 \includegraphics{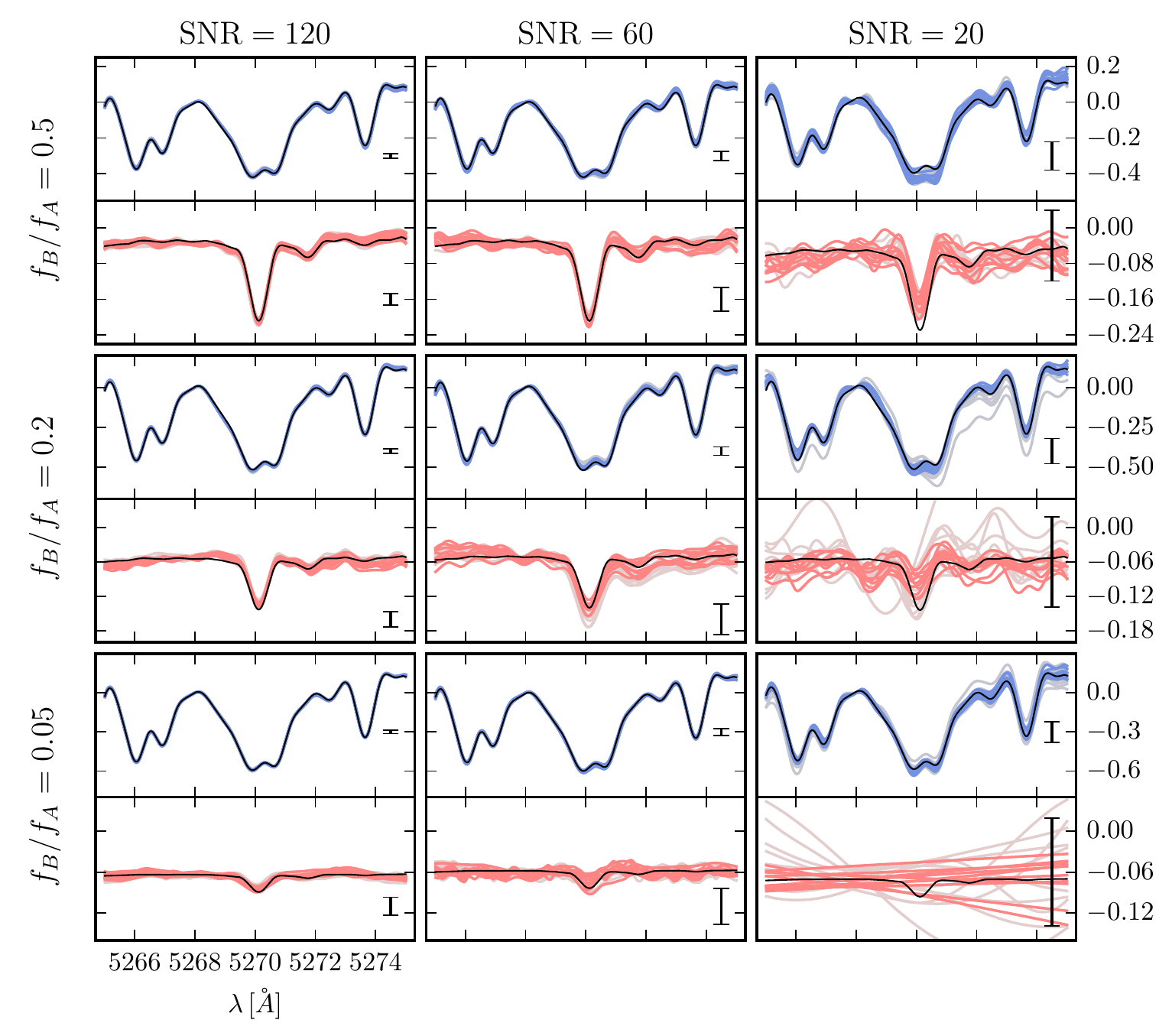}
  \figcaption{Inferred spectra for mock spectroscopic binaries over a grid of signal to noise and flux ratios, with the average noise per pixel is marked in the right of each panel. Note that the vertical scale is different in each row. In each panel, grey lines (background; wider scatter) represent realizations of spectra marginalized over the full orbital posterior. Hued lines (foreground; tighter scatter) represent realizations of spectra generated with the orbital parameters fixed to the fiducial values (i.e., from an external orbital constraint).
  \label{fig:SB2_test_grid}}
  \end{center}
\end{figure}

\begin{figure}[tb]
\begin{center}
 \includegraphics{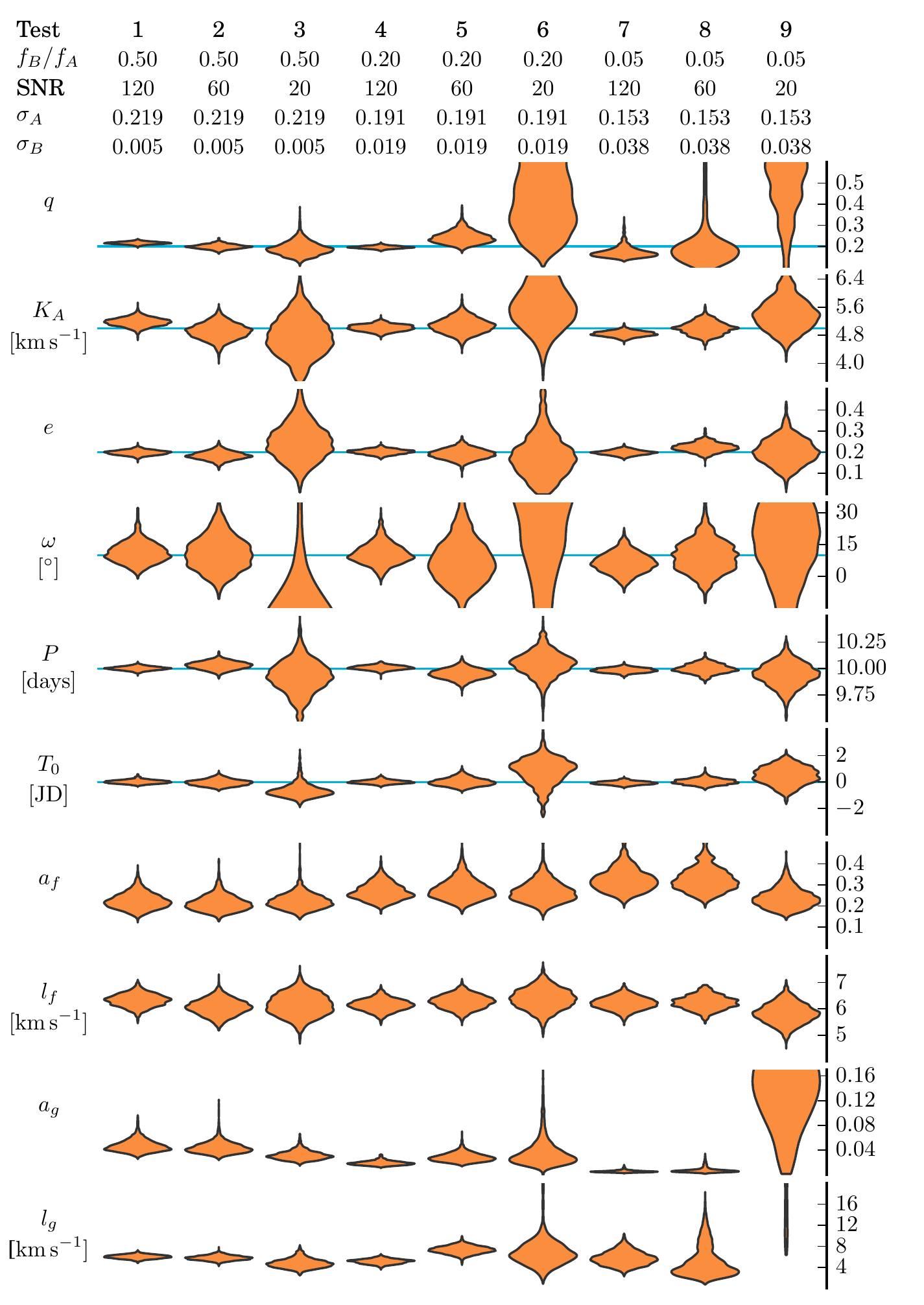}
  \figcaption{Marginal posteriors for the orbital parameters and hyperparameters corresponding to the mock datasets within the test grid. The horizontal blue lines mark the fiducial parameter values, and violin plots denote the one-dimensional marginalized posteriors. The width of the posterior represents the probability density (the widest location of the posterior denotes the most probable parameter value), and posteriors which extend off the range of the plot are generally unconstrained over allowable parameter space.
  \label{fig:SB2_test_grid_posteriors}}
  \end{center}
\end{figure}

Next, we pursue a grid of tests to explore the performance of the Gaussian process model on mock data with a range of flux ratios $f_B/f_A = \{ 0.05, 0.2, 0.5\}$ and signal-to-noise SNR $= \{20, 60, 120\}$. The results of the reconstruction are shown in Figure~\ref{fig:SB2_test_grid}, and a visual representation of the one-dimensional marginal posterior distributions is shown in Figure~\ref{fig:SB2_test_grid_posteriors}. Because of the aforementioned flux ratio ambiguity, we additionally describe our suite of mock data by the standard deviation of each spectrum ($\sigma_A$ and $\sigma_B$), which is an invariant quantity as concerns the performance of the technique.

For some binary systems (e.g., eclipsing systems), it may be possible to precisely constrain the orbital parameters via alternate means (e.g., photometric eclipses) to a higher precision than is otherwise possible with a noisey and/or extreme flux ratio spectroscopic dataset alone.
Therefore in Figure~\ref{fig:SB2_test_grid} we show draws of the inferred spectra marginalized over the full orbital and Gaussian process hyperparameters (pale grey lines, background) as well as the inferred spectra extracted at the fiducial orbital values but varied over the hyperparameters (hued lines, foreground). The input spectra are shown in black, vertically offset by an arbitrary constant to match the Gaussian process draws as before. To reduce the aforementioned vertical scatter of random draws in this plot, we pin the flux level of primary spectrum to be zero at $\lambda = 5268\AA$. The average noise per pixel in the dataset is shown by an error bar in the right of each panel.

Unsurprisingly, the reconstruction technique performs best for high SNR datasets with moderate (near-equal) flux ratios. The primary spectrum is always accurately inferred across our grid of tests. More importantly, the technique can infer secondary spectra for moderately high flux ratios with very noisy data ($f_B/f_A = 0.2$, SNR = 20) and extreme flux ratios with moderate quality data ($f_B/f_A = 0.05$, SNR = 60), albeit with substantial uncertainty. The inferred values of $a_f$ and $a_g$ approximate the values of $\sigma_A$ and $\sigma_B$, while $l_f$ and $l_g$ are a crude measure of the average width of the spectroscopic lines. For the noisiest, most extreme flux ratio test ($f_B/f_A = 0.05$, SNR = 20), the secondary Gaussian process is unable to constrain the shape of the secondary other than showing that the amplitude of its variations must be below a certain level (see Figure~\ref{fig:SB2_test_grid_posteriors}). In this situation, there also emerges a degenerate solution with large $a_g$ and $l_g$: the data spectrum is dominated by the primary star modulo the addition of an approximately flat continuum. This degenerate solution could be avoided by the addition of a prior that favors smaller $l_g$ at the expected width of the secondary lines.
Overall, the results from this grid of tests are extremely encouraging and demonstrate that the Gaussian process framework is able to accurately infer composite spectra to a precision far exceeding the average per-pixel noise in a spectroscopic dataset. It is important to emphasize that in these tests we have only been using a single $10\AA$ chunk of the spectrum; any analyses incorporating a wider swath of spectrum will potentially be much more sensitive.

\subsection{Single-lined Spectroscopic Binaries and Exoplanet Search}
In the limiting case that the flux of the secondary becomes negligible, the system reverts to a single-lined spectroscopic binary and any constraint on the mass ratio ($q$) vanishes, leaving only 5 orbital parameters. If there is a strong belief that light from the primary component dominates the composite spectrum, then one could save computational time and explore a lower-dimensional parameter space with ${\bm \Sigma}_G = {\bm 0}$. However, if there are indeed unrecognized signatures of the secondary in the composite dataset, a constrained single-lined spectroscopic model will deliver biased orbital parameters. When using the full complement of binary orbital parameters, the Gaussian process framework provides the ability to place constraints on the relative flux ratio of an unknown companion in a probabilistic setting. While other search techniques rely upon cross-correlation and subtraction of a presumed template \citep[e.g.,][]{gullikson15}, the Gaussian process framework provides a sensitive, model-free mechanism to detect or place limits on faint spectroscopic lines from a companion star. Of course, secondary contamination is much less an issue in the case of radial velocity measurements of exoplanetary systems around single-stars. While any constraints on the nature of the secondary spectrum might be meaningless, such a Gaussian process technique could be used to infer precision radial velocities for planet search. If a statistically significant non-zero semi-amplitude ($K_A$) and orbital period ($P$) were found for the primary star, this would signal the existence of an orbiting exoplanet. Lastly, the framework could be used to simply infer a high-fidelity template of the primary star, which could be used for abundance analysis or as a template for traditional cross-correlation radial velocity measurements.

In the development of this framework, we have assumed that the intrinsic spectrum of the star
is constant with time, an assumption also made by planet search pipelines using the standard cross-correlation technique.
However, subtle changes in the spectrum due to starspots can bias the inferred radial velocity at the level of $\Delta v \sim 10\,\mathrm{m\, s}^{-1}$ \citep{dumusque11}, dwarfing the operating precision of state of the art spectrographs \citep[$\Delta v \approx 1\,\mathrm{m\, s}^{-1}$;][]{fischer16}.
In the Gaussian process framework, the effects of starspots could be minimized by incorporating an additional time-covariance that allows for flexibility in the exact shape of the spectrum. Such an extension would
be especially useful when simultaneous time-series photometric observations bracket the spectroscopic observations. Measurements of starspot modulation could be used to predict the changes in the spectral line profiles, as is currently done in advanced radial velocity analyses \citep[e.g.,][]{haywood14}.
We further explore the consequences of time variability in Section \ref{subsec:timevariable}.

\section{Results: Application to the Mid-M dwarf binary \obj} \label{sec:results}

\subsection{The \obj\, dataset and traditional orbit analysis}
As part of the MEarth-South survey \citep{irwin15}, \citet{dittmann16} recently discovered the mid-M dwarf \obj\ to be a short-period eclipsing binary.
Subsequent photometric and radial velocity follow-up over two observing seasons constrained the system period to be $P = 4.7043512_{-0.0000010}^{+0.0000013}\,\mathrm{days}$, the primary component mass and radius to be $M_A = 0.30795\pm0.00084\,M_\odot$ and $R_A = 0.3226\pm 0.0033\,R_\odot$, the secondary component mass and radius to be $M_B = 0.19400\pm0.00034\,M_\odot$ and $R_B = 0.2174\pm 0.0023\,R_\odot$, and the distance to be $24.9 \pm 1.3\,\mathrm{pc}$.
Given the lack of strong X-ray emission and the absence of Lithium in the spectra, \obj\ is likely a field-age system. At these masses, both stars are fully-convective \citep[$M< 0.35 M_\odot$;][]{chabrier97}. Such a system provides an interesting application for our spectroscopic disentangling technique because the orbital parameters are known to high accuracy due to the eclipsing nature of the system. For the purposes of demonstrating the disentangling framework, we ignore any eclipse constraints and utilize only the spectroscopic dataset in \citet{dittmann16}. Later, we utilize the additional constraints from the eclipses to discuss the fundamental properties of \obj\ A and B, in particular the stellar radii.

The spectroscopic dataset consists of 14 epochs (1 hour integration each, except for one 40-minute epoch) taken over the course of February 2014 to January 2015 with the Tillinghast Reflector Echelle Spectrograph (\emph{TRES}; $R = \lambda / \Delta \lambda = 44,000$) on the 1.5m Tillinghast reflector at the Fred Lawrence Whipple Observatory on Mt. Hopkins, in Arizona. The data were reduced by the standard \emph{TRES} pipeline \citep{buchhave10}, and placed in the barycentric frame. At red wavelengths, the signal to noise of the spectra ranges from 20 to 60 per resolution element, with a median value of 43. To determine radial velocities with \texttt{TODCOR} \citep{zucker94}, \citet{dittmann16} also obtained high signal-to-noise observations of the following mid-M template stars from \citet{kirkpatrick91}: Gl~273 (M3.5), Gl~699 (Barnard's star; M4), Gl~83.1 (M4.5), and Gl~406 (M6), which we later use for comparison with the disentangled spectra of \obj~A and B.
\citet{dittmann16} found the strongest signal when using the Gl~699 template over the wavelengths $7060\AA - 7200\AA$, which exhibit sharp molecular features, and measured a best-fit component light-ratio of $L_B / L_A = 0.434 \pm 0.025$. Because one or both stars may be spotted and the fraction of eclipsed spots is unknown, \citet{dittmann16} did not quote a measurement from the eclipse measurements. If one relies upon the derived physical parameters for the system, the temperature scale of \citet{mann15}, and the \citet{allard01} models, the predicted light ratio would be $L_B / L_A = 0.406$.
\citet{dittmann16} omitted the observation on HJD~2456743 from their orbital fit because they were unable to separate the primary and secondary peaks of the CCF, meaning that at this epoch each component had a similar radial velocity. Assuming equal measurement uncertainty for the primary and secondary radial velocities, they calculated a best-fit binary orbit with the 13 epochs of measurements and found the average residuals of the primary and secondary velocities to be $0.05\,\kms$ and $0.20\,\kms$, respectively.

\begin{figure}[t]
\begin{center}
 \includegraphics{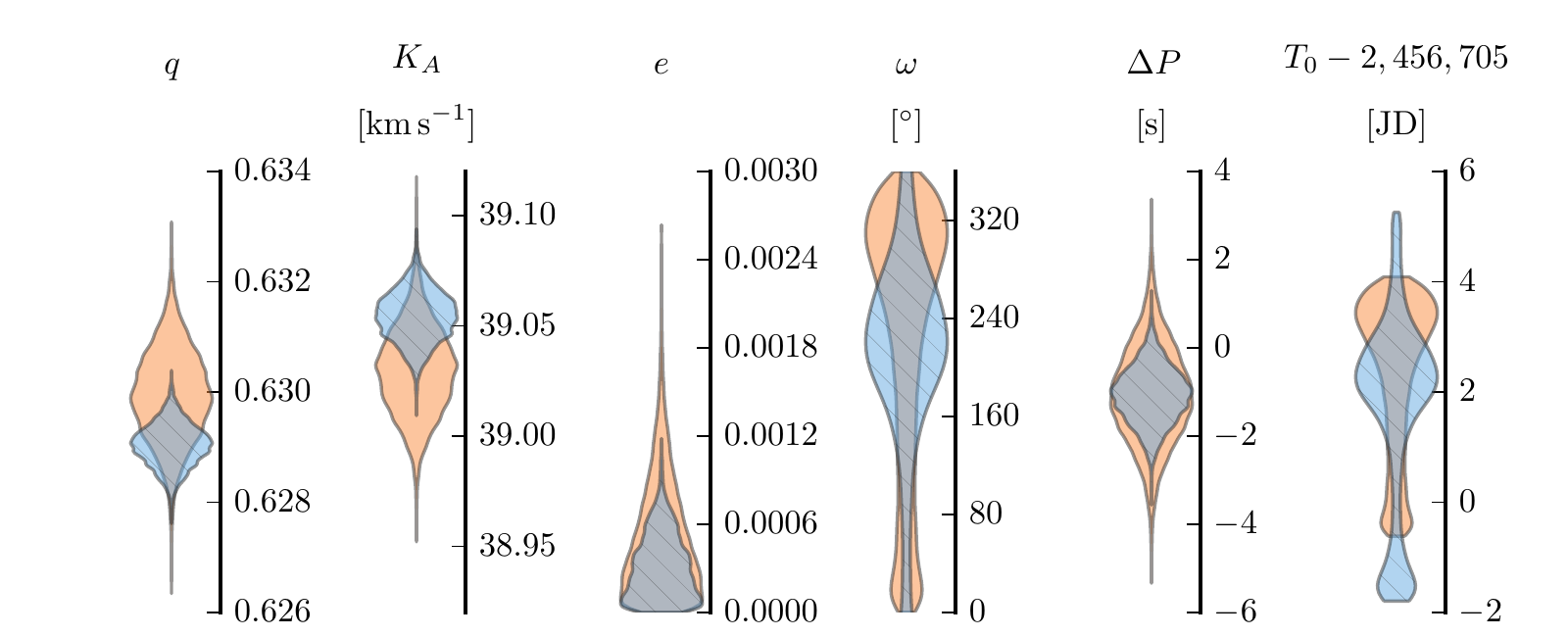}
  \figcaption{The one-dimensional marginal posteriors from the traditional RV analysis (solid orange) and the Gaussian process framework (hatched blue). The $\gamma$ posterior from the traditional analysis is not shown ($\gamma = 0.025 \pm 0.011\,\kms$). The period $P$ is shown relative to the joint photometric and spectroscopic value reported in \citet{dittmann16}, $P =  4.7043512\,\mathrm{days}$.
  \label{fig:rv_comparison}}
  \end{center}
\end{figure}

In order to make a direct comparison with the orbital parameters we will determine with the Gaussian process framework, we re-fit the 13 radial velocities in \citet{dittmann16} using the orbital parameterization described in \S\ref{subsec:orbital_params}. We adopt  $0.05\,\kms$ and $0.20\,\kms$ as the uncertainty on each radial velocity measurement for the primary and secondary, respectively, and employ a simple $\chi^2$ likelihood function with flat prior distributions. We explore the $7$-dimensional posterior with the \texttt{emcee} implementation \citep{foreman-mackey13} of the Affine Invariant Ensemble Sampler \citep{goodman10}. To assess convergence, we run multiple ensembles of 28 walkers for 10,000 iterations each, starting from different locations in parameter space and burning 5,000 iterations worth of samples.
The marginalized, one-dimensional posteriors are shown in Figure~\ref{fig:rv_comparison} and are in excellent agreement with \citet{dittmann16}, except for our determination of $\gamma = 0.025 \pm 0.011\,\kms$. Our measurement is discrepant with the value reported by \citet{dittmann16} ($-0.009 \pm 0.014\,\kms$) at the $2\sigma$ level, which we suspect may be due differences in how the radial velocity uncertainties were incorporated in the fitting process. As noted in \citet{dittmann16}, the orbit is extremely circular, which leads to the common $\omega - T_0$ degeneracy.

In addition to the standard spectral reduction performed by \citet{dittmann16}, we perform additional processing of the spectra to facilitate a uniform comparison between objects. Along with the observations of \obj\ and the 4 aforementioned M-dwarf template stars, we also download an observation of Vega taken on HJD~2456740 from the \emph{TRES} archive (P.I.~D.~Latham) to use as a reference for telluric line contamination. As part of a typical \emph{TRES} observational sequence, observations of a continuum source are taken to characterize the echelle blaze function, which has been shown to be very stable with time for most echelle orders. We divide each spectrum by its associated blaze measurement to produce approximately flat spectra. This process does not directly lead to flux-calibrated spectra, however, since the absolute brightness of the source is not known. Nor does this process necessarily lead to spectra that match the same shape as flux-calibrated spectra, since the spectrum of the continuum source may be significantly different than that of a source originating from above the atmosphere. A \emph{TRES} calibration program (P.I.~I.~Czekala) repeatedly observed several spectrophotometric standard stars over several months and demonstrated that the \emph{TRES} intra-order throughput shape is very stable with time. Since the bandpass sensitivity in each order is well-described by a 4th-order Chebyshev polynomial, this means that from night to night the linear and higher order coefficients varied less than $4\%$. The inter-order sensitivity is less stable, with nightly scatter typically $\lesssim 10\%$.

Large scale, sloping features of M-dwarf spectra are important for spectral typing and characterization. Although we do not have observations of spectrophotometric standards contemporaneous with the \obj\ observations, because the intra-order \emph{TRES} throughput is stable we can approximate flux-calibrated \obj\ spectra by applying the sensitivity functions derived from the calibration program, which were measured from  a series of observations of the O2 spectrophotometric standard BD+28~4211 in November 2013. Because the inter-order flux calibrations are not as stable as the intra-order calibrations, we further rectify the spectra so that the median flux of each order is 1. The result of this process is spectra whose intra-order shape calibration should be comparable to a bona fide flux calibration within $4\%$. Because of the per-order rectification process, the inter-order shape calibration will not be correct. It should be noted that if same-night spectrophotometric flux observations existed, the proper shape of the entire spectrum could be recovered.


\subsection{Gaussian process inference of orbital parameters \\ and the disentangled spectra of A and B}

\begin{figure}[t]
\begin{center}
 \includegraphics{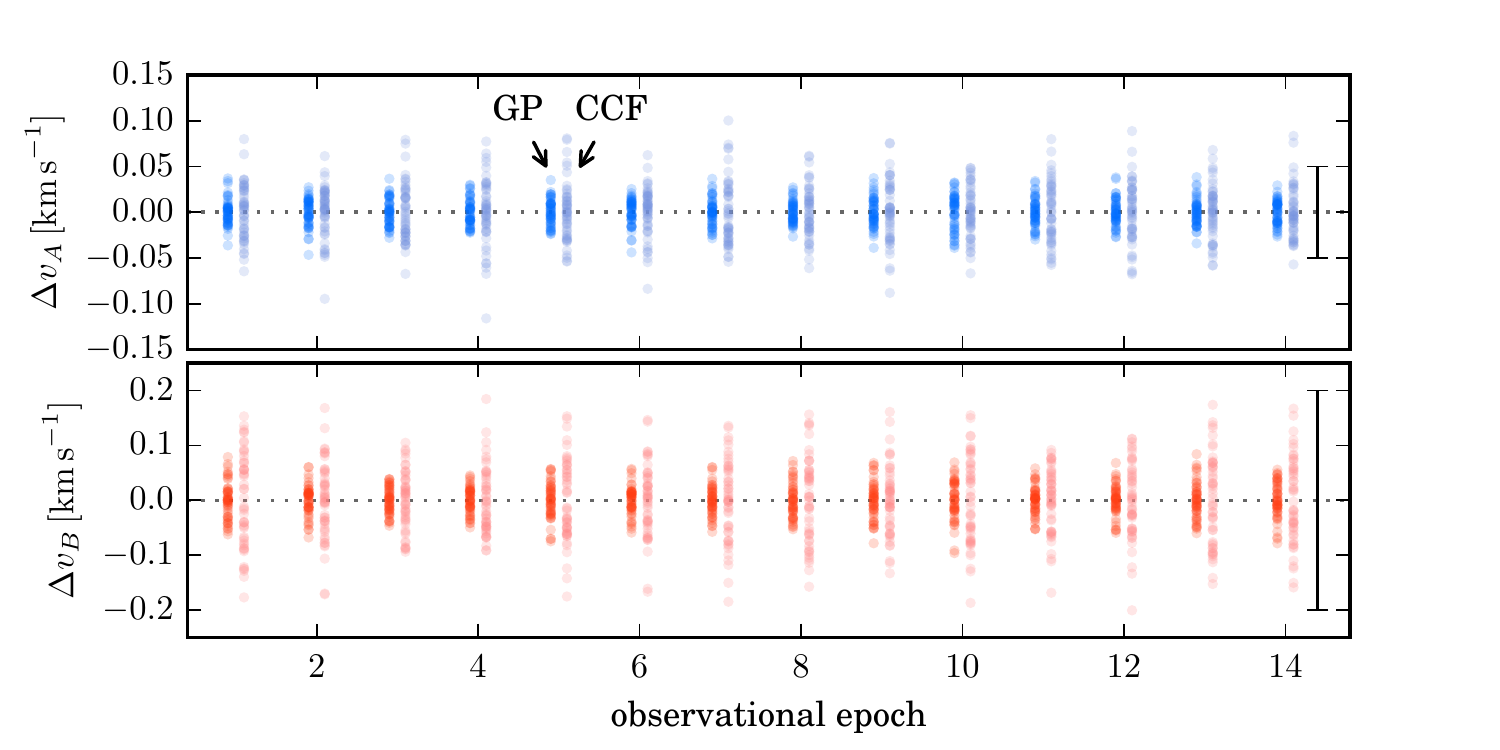}
  \figcaption{The relative radial velocity precision of the Gaussian process based orbit (GP, left, darker points) compared to the traditional cross-correlation based orbit (CCF, right, lighter points) on an epoch-to-epoch basis. For each technique, 50 sets of orbital parameters are drawn from the posterior probability distribution (Figure~\ref{fig:rv_comparison}), radial velocities are computed for each observational epoch, and then are plotted relative to the mean radial velocity at that epoch. The CCF radial velocity uncertainty measured by \citet{dittmann16} is shown by the errorbar in the right of each panel.
  The per-epoch scatter of the Gaussian process technique delivers a per-epoch radial velocity scatter that is $\sim 50\%$ of that of the traditional CCF technique.
  \label{fig:rv_scatter}}
  \end{center}
\end{figure}

We apply our spectroscopic binary Gaussian process framework to the pseudo-flux-calibrated dataset of \obj\ in a similar manner as in \S\ref{subsec:mock}. For computational efficiency, we focus on two high signal to noise regions of the spectrum uncontaminated by telluric lines, $6650\AA - 6770\AA$ and $7060\AA - 7140\AA$. Unlike with traditional cross-correlation radial velocity analysis, composite spectra with near-equal component radial velocities still provide useful information in our Gaussian process framework and help constrain the underlying intrinsic component spectra. Therefore we use all 14 epochs of spectroscopy in our analysis, including the epoch omitted by \citet{dittmann16}. We divide the dataset into 20 chunks of $10\AA$ each so that the evaluation of the posterior function can be parallelized across a computer cluster. The posterior distribution for each $10\AA$ chunk is evaluated on an individual core, and then gathered and multiplied together to yield a final posterior for the full spectral range. This final posterior is explored using a simple Metropolis-Hastings MCMC,\footnote{included in \texttt{emcee}.} whose jump covariances have been tuned to the structure of the posterior for efficient exploration of the parameter space. We run multiple chains from different starting locations for 20,000 iterations each and assess their convergence by the Gelman-Rubin statistic $\hat{R}$ \citep{gelman13}.
The one-dimensional marginalized posteriors are shown in Figure~\ref{fig:rv_comparison} in blue. We recover similar posteriors as the traditional cross-correlation analysis, and notably several parameters are obtained with higher precision (e.g., $q$, $K_A$, $e$, and $P$). With more computational resources, a larger spectral range could be included in the analysis and potentially tighten the parameter constraints even further. In Figure~\ref{fig:rv_scatter} we further illustrate the increased precision of the Gaussian process based technique by demonstrating that it delivers a smaller radial velocity scatter at each observational epoch compared to the traditional cross-correlation based technique.

\begin{figure}[htb]
\begin{center}
 \includegraphics{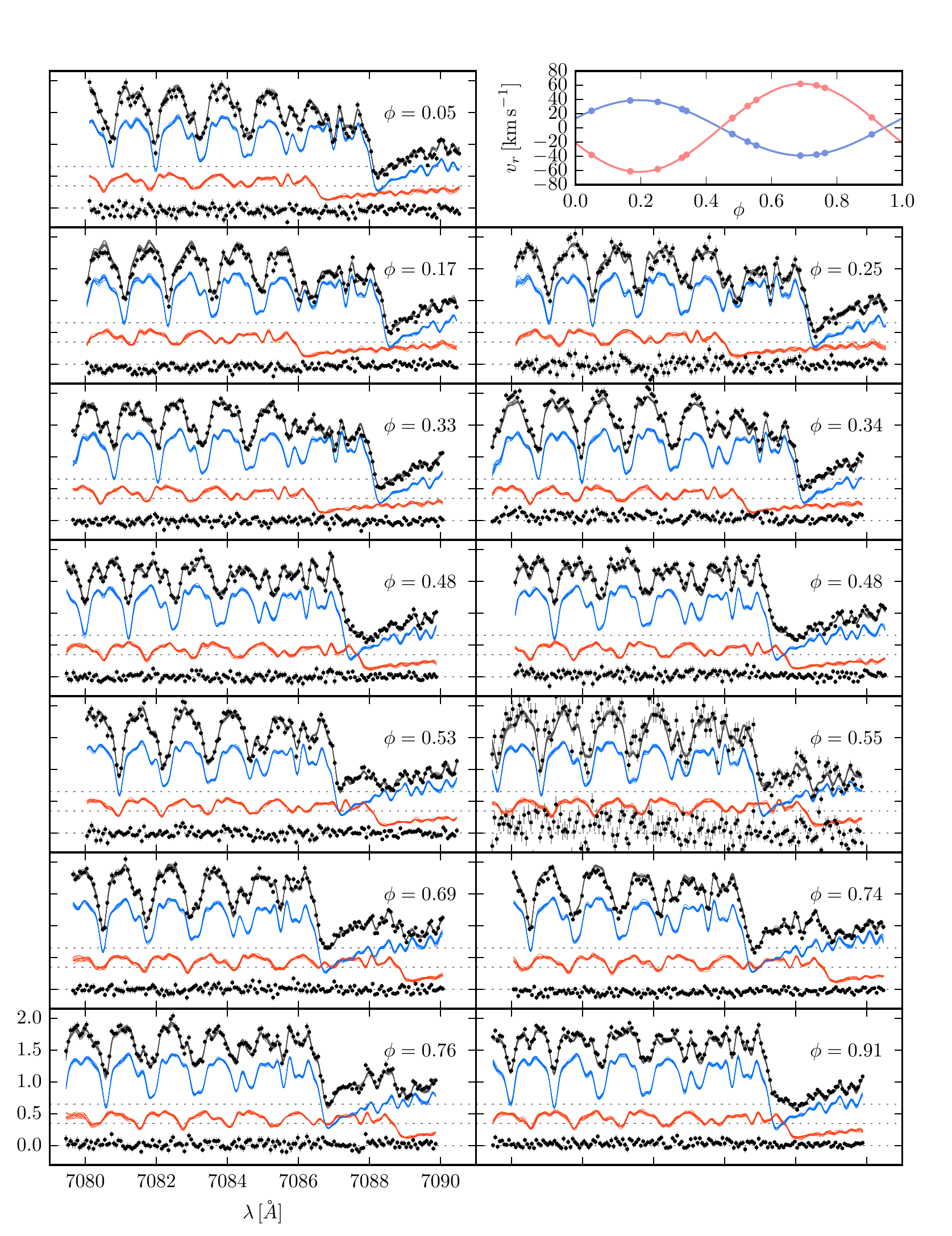}
  \figcaption{Thirteen epochs of composite spectra are shown for a highlighted $10\AA$ chunk. A fourteenth epoch at phase $\phi=0.03$ was omitted from this figure for space reasons, since it is the same phase as the first plot but lower signal-to-noise. 10 realizations of the primary and secondary are drawn from the posterior predictive distribution (blue and red, respectively), and residuals are computed from the mean prediction.
  \label{fig:LP611_all_epoch}}
  \end{center}
\end{figure}

An example of the analysis for a single chunk of \obj\ spectrum is shown in Figure~\ref{fig:LP611_all_epoch}, focusing on a particularly dramatic bandhead for emphasis. The composite dataset for each epoch is shown in black, with the corresponding orbital phase labeled in the upper right. Draws of the Gaussian process are generated while varying over random draws of the orbital parameters and hyperparameters from the posterior probability distribution, showing that the distribution of reconstructed component spectra fit well within the noise. As noted in \S\ref{subsec:mock}, there is the additional question of choosing the normalization level for each reconstructed spectrum. And so in Figure~\ref{fig:LP611_all_epoch}, we arbitrarily choose ${\bm \mu}_f = 0.65$ and ${\bm \mu}_g = 0.35$ to display these spectra (denoted by horizontal dashed lines). We also show the residuals computed from the mean prediction of the composite spectrum at each epoch. Because the spectra have a natural covariance with wavelength, any individual realization of residuals may show a slight covariance as well.

\begin{figure}[tb]
\begin{center}
 \includegraphics{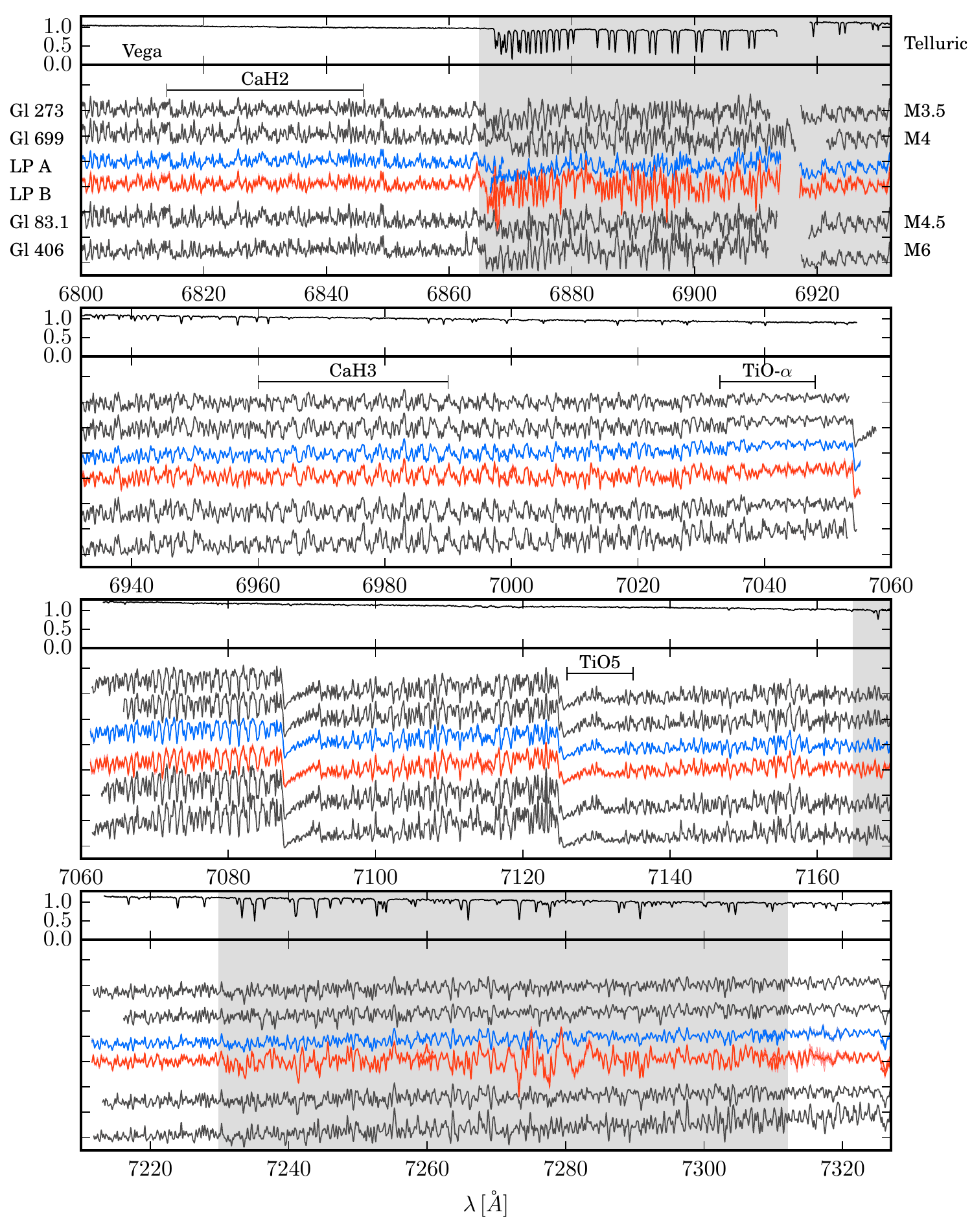}
  \figcaption{The reconstruction of \obj~A and B, showing 10 realizations of the spectra (thin, faint lines) overplotted with the mean prediction (thick, hued lines), compared to four mid-M spectral templates in black: Gl~273 (M3.5), Gl~699 (Barnard's star; M4), Gl~83.1 (M4.5), and Gl~406 (M6). Common spectral indices used to type M-dwarfs are marked: CaH2, CaH3, TiO-$\alpha$, and TiO5. A spectrum of Vega is shown on top to highlight areas contaminated by telluric lines. Severely disrupted regions are marked with a grey background.
  \label{fig:LP611_comparison_blue}}
  \end{center}
\end{figure}

\begin{figure}[tb]
\begin{center}
 \includegraphics{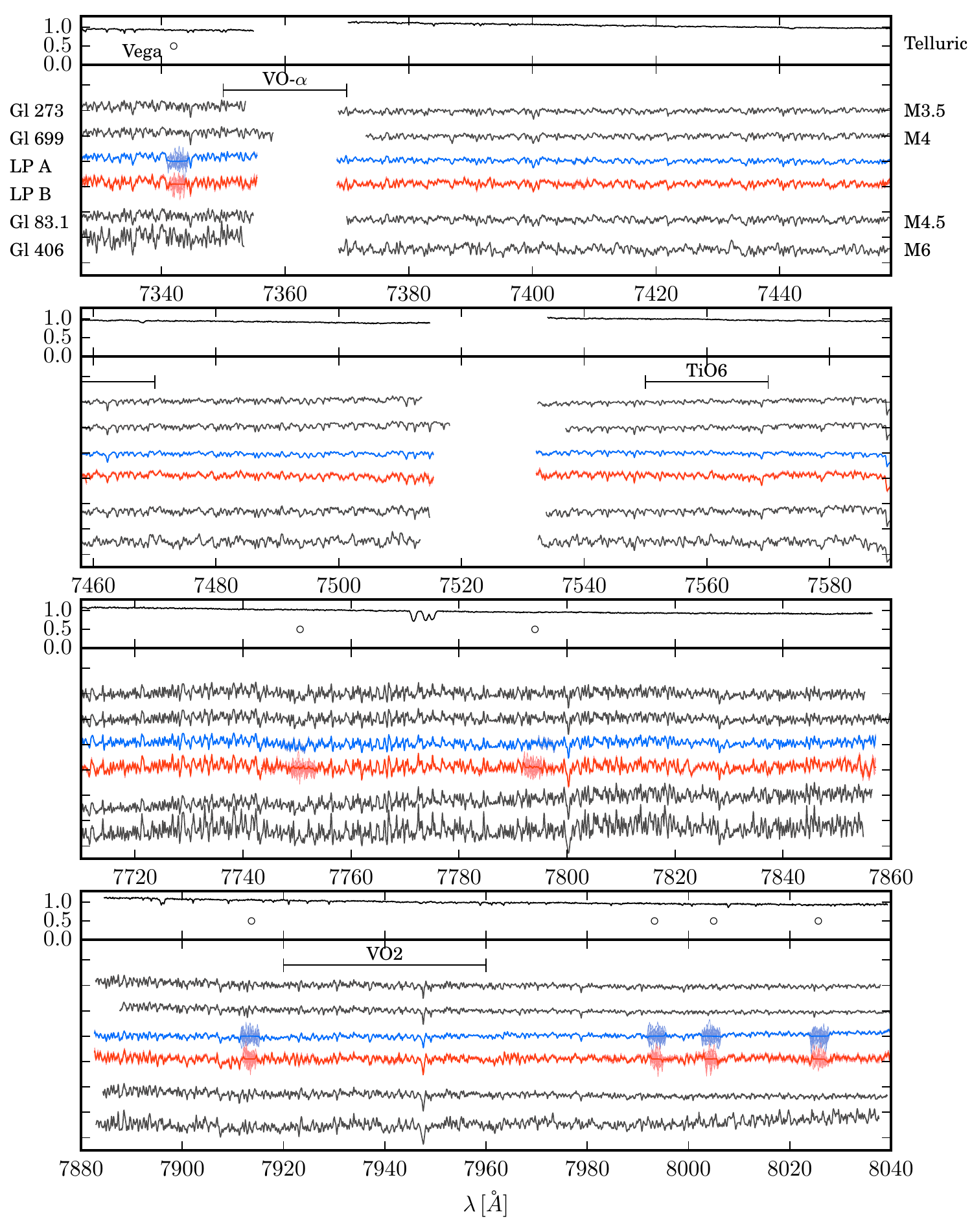}
  \figcaption{A continuation of Figure~\ref{fig:LP611_comparison_blue}. Regions that were masked due to poor subtraction of night sky emission lines are marked with a $\circ$ symbol. Because these regions are relatively unconstrained by data, the reconstructed spectra revert to draws from the Gaussian process prior.
  \label{fig:LP611_comparison_red}}
  \end{center}
\end{figure}

While the exploration of the posterior is performed in $10\AA$ chunks and using only a subset of the full spectral range, by assuming the inferred orbital parameters and Gaussian process hyperparameters are valid outside of this range, one can reconstruct a wide range of component spectra. In Figures~\ref{fig:LP611_comparison_blue} and \ref{fig:LP611_comparison_red} we show the reconstructed spectra over the full wavelength range $6800\AA - 8040\AA$, which includes many of the notable spectral regions used for classifying M-dwarfs. We also overplot an observation of Vega, a fast rotating A0 star with a near-featureless continuum, in order to highlight regions affected by telluric lines. Any spectral features intrinsic to Vega (e.g., \ion{O}{1} $\lambda \lambda \lambda 7771$) will be broadened by its fast rotational velocity \citep[$v \sin i = 21.9\,\kms$;][]{hill04}, and contrast with the narrower telluric lines.
In contrast to the carefully chosen narrow range we used for the orbital parameter inference, this expanded region includes numerous telluric absorption lines and night sky emission lines. Because these features are not consistent with the rest-frame of either star A or B, they act to distort the reconstruction process by acting as outlier flux points that cannot be reproduced by the model. In principle, it should be possible to model the effects of telluric lines as an additional Gaussian process since their location is known precisely, although their absorption depth changes from epoch to epoch with airmass and atmospheric conditions. Simultaneous modeling of the telluric features would allow us to access spectral information of the stars in regions which are currently contaminated. We note that other spectral disentangling methods have successfully modeled telluric lines \citep{hadrava06}, but we leave a detailed implementation of these concepts for future work.

Typically, night sky emission lines are subtracted by the \emph{TRES} reduction pipeline, although in some instances the subtraction is sub-optimal and leaves residual flux. We denote regions affected by poor-night sky line subtraction by a $\circ$ symbol and mask these features in the dataset \citep{osterbrock96}.
Because $\sim 5 \AA$ windows surrounding these lines have been masked for all epochs, there are regions for which $f$ and $g$ are minimally constrained. In these regions, the Gaussian process mean is generally a flat line, and the draws mimic draws from the prior distribution. Were we to predict intrinsic spectra outside of the range of the detector (e.g., in the coverage gaps between echelle orders), a similar phenomenon would occur.

To properly plot the reconstructed \obj~A and B spectra (i.e., draws of the mean ${\bm 0}$ Gaussian processes conditional on the composite data) in a format suitable for comparison with other rectified observations of mid-M template stars, we must first renormalize the spectra. Renormalization refers to the process whereby the spectra are brought to their correct continuum level in the frame of the data with the addition of a constant, and then rectified to a mean of 1 by dividing by that same constant
\begin{equation}
\hat{{\bm F}} = \frac{{\bm F} + {\bm \mu}_f}{{\bm \mu}_f} \quad\quad  \hat{{\bm G}} = \frac{{\bm G} + {\bm \mu}_g}{{\bm \mu}_g}.
\label{eqn:rectify}
\end{equation}
For \obj~A and B, we assume that the mean vectors are simply flat with wavelength, but in principle their relative amplitude could be guided by beliefs about how the flux ratio changes with wavelength, (i.e., ${\bm \mu}_f(\lambda)$, and ${\bm \mu}_g(\lambda)$). An incorrect renormalization will distort the continuum level and the contrast of the spectral lines. Because the secondary spectra are intrinsically fainter, they must be scaled by a larger constant to be rectified, which has the effect of making the rectified spectrum appear noisier.
We experiment varying the flux ratio until we find renormalizations of \obj~A and B that match the amplitude and slope of the nearby mid-M templates. We find that a value of ${\bm \mu}_g/{\bm \mu}_f = 0.33$ yields a renormalization that has the spectra of \obj~A and B closely matching the mid-M templates over the full wavelength range. We note that flux ratios as high as 0.434 \citep{dittmann16} yield acceptable results over much of the spectral range, however the contrast of the bandhead at $7090\AA$ appears too shallow in the secondary spectrum. \citet{fremat05} discuss an alternate renormalization strategy suitable for stars with very deep absorption lines, which must remain positive even in their deep cores, and so they set a lower limit on the continuum level of the spectrum. Unfortunately, no such lines exist in the \obj\ spectra.

\subsection{\obj~A and B in the mid M-dwarf context}

Now, with the disentangled spectra of \obj~A and B known (up to a modest uncertainty in the renormalization factor), we can examine the stars in the context of the other mid-M spectral templates. Although we have assumed that the flux ratio between A and B is constant across the full wavelength range, given that both stars are evidently very close in spectral type this appears to have been a reasonable decision. Next to \obj~A and B, we also plot \emph{TRES} observations of mid-M standard stars \citep{kirkpatrick91} and highlight several spectral regions commonly used in classification, which generally include molecular features. With low to moderate resolution spectra, the flux within these bands measured relative to a nearby pseudo-continuum region serves as an index for objective spectral typing. Some of the most widely indices used are summarized by \citet{lepine13}: CaH2, CaH3, and TiO5 \citep{reid95}; VO1 \citep{hawley02}; and TiO6 and VO2 \citep{lepine03}. Those most sensitive (i.e., changing most rapidly) for M3 - M6 stars are the TiO5, TiO6, and VO2 indices \citep{lepine13}.

We would like to directly measure the indices for the disentangled spectra of \obj~A and B and report an index-based spectral type for these stars; however, when we measured the indices of the template stars in common with \citet{lepine13}, we were unable to reproduce their values. Since we were able to recover the general trends with spectral type, we suspect that this discrepancy is likely due to the higher resolution of our instrument and the fact that our spectra are not flux-calibrated following a rigorous procedure. Therefore instead of comparing indices, we directly compare the morphology of the spectra within the noted index regions. We find that \obj~A and B naturally fall in a sequence of mid-M standards near Gl~699 (M4) and Gl83.1 (M4.5), which agrees well with the previously determined composite spectral types of M3.5 \citep[optical;][]{reid04} and M4.27 \citep[NIR;][]{terrien12}.

We emphasize that other than the procedure performed in Equation~\ref{eqn:rectify} for \obj~A and B, no post-processing (e.g., continuum or pseudo-continuum normalization) has been applied to any of the spectra presented in Figures~\ref{fig:LP611_comparison_blue} and \ref{fig:LP611_comparison_red}. If a proper flux-calibration of the data were performed on the raw data and the flux ratio correctly specified, then the inferred shape of the reconstructed spectra would be correct. The ability of the Gaussian process technique to constrain the low order shape of inferred spectra avoids a potential downside of Fourier disentangling techniques.

The Gaussian process technique can also disentangle emission lines, and we find that both \obj~A and B show strong Balmer series emission---in fact, the radial velocity semi-amplitude is large enough such that the emission lines from A and B are completely separated at quadrature. The strong Balmer emission indicates that both stars are magnetically and chromospherically active \citep{hawley96,west11,lepine13}, possibly as a result of rapid, synchronous rotation due to tidal locking \citep{dittmann16,newton17}. Because magnetic activity serves to trap flux and inflate stellar radii while decreasing effective temperature \citep{morales09,morales10}, it may provide an explanation for the inflated radii of \obj~A and B measured by \citet{dittmann16}.

\section{Discussion} \label{sec:discussion}

\subsection{Hierarchical inference of spectral chunks} \label{subsec:hierarchical}
In our spectroscopic binary Gaussian process framework, we made the assumption that the Gaussian process kernels were the same functional form for each spectrum, and that each $10\AA$ chunk of spectrum used the same Gaussian process hyperparameters. If the flux ratios of the spectra are approximately constant over the chosen wavelength range, and the relative amplitudes of the spectral line variations are similar, as in \obj, then this approach is adequate. However, if we are inferring a large region of the spectrum which might contain spectral features of very different morphologies, then a more accurate spectral extraction might be obtained by modeling the Gaussian process hierarchically, where the amplitude and length scale for an individual chunk is drawn from a general hyperparameter distribution. For example, the amplitude ratio could be guided by the ratio of fluxes expected from the hypothesized effective temperatures, with the amplitude of each individual chunk allowed to vary somewhat around this mean value.
Recently, high contrast observational techniques have shown great promise to detect the signatures of exoplanet atmosphere spectra via cross-correlation with templates \citep[e.g.,][]{snellen14,birkby17}--with increasing telescope sensitivity and an appropriately flexible Gaussian process model it may become possible to reconstruct the exoplanet spectra directly.

Although we have only dealt with a single dataset acquired by the same instrument in an identical setup, in principle, data from multiple instruments covering different wavelength ranges (e.g., optical and infrared spectra) could be inferred simultaneously if the Gaussian process hyperparameters were allowed to vary individually for each instrument, tracking changes in the line spread function. Moreover, since the convolution of a Gaussian process with a line spread function is also a Gaussian process, it may even be possible to simultaneously model data taken with different instrumental setups but covering the same wavelength range if the latent spectrum is modeled at high resolution and convolved down to the resolution of each observation.

\subsection{Extensions to time-variable spectra} \label{subsec:timevariable}

Thus far, we have only discussed covariances with a $\lambda$-dependence. If one or more of the components is believed to be time-variable, for example due to chromospheric activity, then we could introduce an additional $t$-dependence to the covariance kernel as well. Until now, we have essentially enforced that the Gaussian process has an infinite covariance with time: pixels with nearby wavelengths contribute equally to constraining the stellar flux, regardless of measurement epoch. If we had an additional time-variable component, then we could allow subsequent epochs to vary such that only flux measurements that were close in wavelength \emph{and} close in time would constrain the inferred spectrum. This also means that we would be able to make predictions (realizations) of the stellar spectrum not only as a function of wavelength, but also as a function of time. Such a mechanism could provide radial velocity measurements for the noiseiest stars, where stellar activity has hampered analysis techniques more profitably applied to quiet stars. In particular, probing the youngest exoplanet population has been hindered by stellar activity, requiring extensive analysis to prove that a planetary signal is not simply unexplained variability \citep[e.g.,][]{johns-krull16}. For more mainstream exoplanet radial velocity analysis, a time-variable, $\lambda$-specific kernel could provide a natural way to identify or downweight spectral regions which bias a traditional radial velocity signal. It may also be possible to adapt the Gaussian process framework to model the iodine absorption lines commonly imprinted on a stellar spectrum for use in precision radial velocity analysis.

While we have focused on inferring stellar photospheres, the spectrum to be inferred need not necessarily be stellar in origin. A potentially exciting application is to use a Gaussian process to infer active signs of accretion in young T~Tauri stars. While the choice of squared exponential kernel has worked well for modeling the intrinsic stellar spectrum of \obj, for other types of stars or variable phenomena, different kernels might be more effective. For example, using the \matern\ kernel for the $t$-correlation might provide a Gaussian process that more closely reflects the stochastic accretion process. Such a modeling approach could also be combined with a mean function ${\bm \mu}_f(\lambda)$ specified by flexible physics-based models \citep[e.g.,][]{czekala15b,gully-santiago17} in order to simultaneously determine the fundamental properties of the veiled star, a traditionally difficult process \citep{herczeg14}.

\section{Summary} \label{sec:summary}

We have demonstrated the successful application of Gaussian processes for inferring component spectra of single and double-lined spectroscopic binaries, while simultaneously exploring the posteriors of the orbital parameters and the spectra themselves. We evaluated the performance of the technique by using mock data constructed over a range of flux and signal to noise ratios, and demonstrated that the technique has desirable characteristics for a wide range of data qualities, returning unbiased posterior estimates. In contrast to Fourier-domain methods, our technique excels at reconstructing the low-order shape of the component spectra and provides a natural and well-motivated probabilistic formalism to model the data in its natural pixel space. While other $\lambda$-based techniques do exist \citep[e.g.,][]{simon94}, our Gaussian process formalism can account for the natural $\lambda$-covariances in each spectrum, providing a natural ``de-noising'' of the spectra typically offered by Fourier techniques.

We applied our framework to the mid-M eclipsing binary \obj. Using only a limited number of observations with modest SNR, we simultaneously recovered both the A and B components and inferred the orbital parameters with a precision exceeding that of standard cross-correlation techniques. For the first time, we independently analyzed each component of \obj\ in the context of a mid-M spectral sequence, and determined that both components are M4 systems, with A appearing slightly earlier than B. We discussed potential applications of Gaussian processes for spectral modeling, including precision exoplanet radial velocity work around active young stars. The code used in our analysis is available as a freely available open-source package.\footnote{Precision Spectroscopic Orbits A-Parametrically (PSOAP) available at \url{https://github.com/iancze/PSOAP}}

\acknowledgments
IC would like to thank the following people:
Guillermo Torres and Dave Latham for advice and helpful discussions about fundamental stellar properties, stellar orbits, and radial velocity techniques;
Kevin Gullikson for guidance about searching time series spectra for faint companions;
Hyungsuk Tak for discussions about Quasar light curve analysis with Gaussian processes;
Kevin Hardegree-Ullman for references to M-dwarf spectral indices;
Eric Nielsen for discussions about orbital dynamics;
Greg Green and Dan Foreman-Mackey for general discussions of Gaussian processes;
Charlie Conroy and the hosts of the ``Fitting Stars, CMDs, and Galaxies'' workshop for insightful discussion about stellar inference than influenced this work;
and Allyson Bieryla, Gil Escuerdo, and Jessica Mink for their service in support of the \emph{TRES} instrument. KM is supported at Harvard by NSF grants AST-1211196 and AST-156854.
Work by BTM was performed under contract with the Jet Propulsion Laboratory (JPL) funded by NASA through the Sagan Fellowship Program executed by the NASA Exoplanet Science Institute. This material was based upon work partially supported by the National Science Foundation under Grant DMS-1127914 to the Statistical and Applied Mathematical Sciences Institute. Any opinions, findings, and conclusions or recommendations expressed in this material are those of the authors and do not necessarily reflect the views of the National Science Foundation. 

\vspace{5mm}
\facilities{FLWO:1.5m/TRES}

\software{Python, astropy, PSOAP}

\bibliographystyle{aasjournal.bst}
\bibliography{GPspec.bib}


\end{document}